\documentclass[conference]{IEEEtran}

\usepackage{cite}
\usepackage{amsmath,amssymb,amsfonts}
\usepackage{algorithmic}
\usepackage{graphicx}
\usepackage{textcomp}
\usepackage{balance}

\def\BibTeX{{\rm B\kern-.05em{\sc i\kern-.025em b}\kern-.08em
    T\kern-.1667em\lower.7ex\hbox{E}\kern-.125emX}}

\begin{document}

\title{Predicting the Performance-Cost Trade-off of Applications Across Multiple Systems}

\author{
\IEEEauthorblockN{
    Amir Nassereldine\textsuperscript{1},
    Safaa Diab\textsuperscript{1},
    Mohammed Baydoun\textsuperscript{1},
    Kenneth Leach\textsuperscript{2},
    Maxim Alt\textsuperscript{2},
    \\
    Dejan Milojicic\textsuperscript{2},
    Izzat El Hajj\textsuperscript{1}
}
\IEEEauthorblockA{\\\textit{\textsuperscript{1}American University of Beirut, Beirut, Lebanon ~~~ \textsuperscript{2}Hewlett Packard Enterprise, Milpitas, CA, USA} \\
}
}

\maketitle

\pagestyle{plain}

\begin{abstract}

In modern computing environments, users may have multiple systems accessible to them such as local clusters, private clouds, or public clouds.
This abundance of choices makes it difficult for users to select the system and configuration for running an application that best meet their performance and cost objectives.
To assist such users, we propose a prediction tool that predicts the full performance-cost trade-off space of an application across multiple systems.
Our tool runs and profiles a submitted application on a small number of configurations from some of the systems, and uses that information to predict the application's performance on all configurations in all systems.
The prediction models are trained offline with data collected from running a large number of applications on a wide variety of configurations.
Notable aspects of our tool include: providing different scopes of prediction with varying online profiling requirements, automating the selection of the small number of configurations and systems used for online profiling, performing online profiling using partial runs thereby make predictions for applications without running them to completion, employing a classifier to distinguish applications that scale well from those that scale poorly, and predicting the sensitivity of applications to interference from other users.
We evaluate our tool using 69 data analytics and scientific computing benchmarks executing on three different single-node CPU systems with 8-9 configurations each and show that it can achieve low prediction error with modest profiling overhead.

\end{abstract}


\section{Introduction}

Application developers today have multiple systems accessible to them for running their applications.
They can choose between different local clusters or private clouds available at their company or institution, as well as a wide range of public cloud offerings.
However, selecting the best system and configuration for an application is challenging because it depends on the behavior of the application as well as the performance and cost objectives of the user.
On the application side, different applications have different performance bottlenecks causing them to perform best on different systems.
Moreover, when applications are given more resources on a system, some may scale well experiencing little increase in cost, others may suffer from increasing cost with diminishing performance returns, and others may even slow down increasing both execution time and cost.
On the user side, some users may aim to minimize cost, others may aim to minimize execution time, and others may aim to find the best trade-off between the two.

Our goal is to provide users with a complete view of the performance-cost trade-off space for any application across multiple systems and configurations.
This information can assist users with making the best decisions that meet their performance and cost objectives.
However, running an application on all systems and configurations to provide such information would be prohibitively expensive.
For this reason, it is desirable to have performance modeling tools that can predict the performance-cost trade-off for arbitrary applications without the need to run these applications to completion and on a large number of systems and configurations.

Many prior works aim at assisting users with selecting the best system and configuration for an application by employing prediction techniques.
One major direction is to use optimization techniques to find the optimal or near-optimal system and configuration that minimize some objective function~\cite{cherrypick,hsu2018arrow,liu2019accordia,klimovic2018selecta,hsu2018micky,xiao2019sara,vanir,casimiro2020lynceus,bilal2020best,hsu2018scout,lin2020fife,wu2021apollo}.
However, simply finding a near-optimal configuration does not give users enough flexibility in meeting their diverse goals and preferences.
For example, a user may be willing to sacrifice a small amount of performance in return for substantial cost savings.
Such decisions can best be made when the user has a full view of the performance-cost trade-off space.

A number of prior works aim at predicting the performance of applications on all available systems and configurations~\cite{ernest,baughman2018profiling,aaziz2018modeling,paris,xu2020arvmec,mariani2017predicting,paragon,quasar,lin2020modelling,chen2021silhouette}.
One approach is to use analytical models~\cite{ernest,baughman2018profiling,aaziz2018modeling}, and train the coefficients of these analytical models for each system using data collected about the application.
However, these models tend to be specific to the application and system in question and do not predict relative performance across multiple systems.
Moreover, they tend to require running the application to completion many times on a system to make predictions for that system.
Another approach is to leverage information collected about other applications offline to train a general model that can make predictions for new applications with little online data collection~\cite{paris,xu2020arvmec,mariani2017predicting,paragon,quasar}.
This latter approach is the one adopted in our work.

We propose a tool for predicting the performance-cost trade-off of an arbitrary application across different systems and different resource configurations per system.
Our tool profiles a submitted application on a small number of configurations called \textit{fingerprint configurations} to generate an application fingerprint.
It then passes the fingerprint to a classifier that distinguishes applications that scale well from those that scale poorly.
For applications that scale well, it passes the fingerprint to a regression model that predicts the performance of the application on all systems and for all resource configurations in each system.
For applications that scale poorly, it uses a different regression model that only predicts the performance of the application on the smallest resource configuration of each system.
Our tool also predicts the sensitivity of the application to different kinds of interference.
To train our prediction models, we run a large number of benchmarks offline on different configurations and collect performance and profiling data.
Using this data, we automatically identify the best set of fingerprint configurations and profiling metrics to use for generating fingerprints, and train the classifier and the regression models accordingly.

Our general approach of generating application fingerprints using profiling information is inspired by the work in PARIS~\cite{paris}.
However, we make several additional contributions.
We provide multiple scopes of prediction (global, single-system, and local) with varying accuracy and online profiling overhead.
We automate the process of selecting the fingerprint configurations rather than hand-picking them.
We make predictions based on relative metrics collected from partial runs without needing to measure execution time from complete runs, and therefore do not require the user to provide a representative short-running task for the fingerprinting process.
We use a classifier to distinguish applications that scale well from those that scale poorly to allow the regression models to focus on a particular class of applications.
We predict the sensitivity of the application to different kinds of interference.
We also quantify the impact of not performing exhaustive data collection when gathering training data offline.

We evaluate our tool using 69 data analytics and scientific computing benchmarks executing on three different single-node CPU systems with 8-9 configurations each (26 configurations in total).
Our evaluation shows that our tool can predict the performance-cost trade-off of an application across these three systems (26 configurations) with a mean error of 22.5\% by profiling the application for just 30~seconds on only three configurations.
Moreover, if we narrow the scope to a single system, our tool can make predictions with a mean error of 11.4\% to 15.6\%.
If we further narrow the scope, our tool can profile applications on a single configuration and make predictions for nearby configurations with less than 10\% error for most configurations.
We show that our tool is effective in realistic scenarios by predicting the performance-cost trade-off of GROMACS~\cite{van2005gromacs}, a commonly used molecular dynamics simulation application, across three systems with a mean error 17.3\% by profiling it for only 5\% of its execution time.
We show that our proposed classifier is responsible for reducing prediction error by a mean of 6.67\%.
We also quantify the impact of making predictions for applications based on partial runs instead of complete runs, and of training our models with partial training data coverage instead of full coverage.
Our future work involves expanding the scope of our tool and evaluation to a wider variety of systems, including multi-node systems with network communication and systems with accelerators such as GPUs.

The rest of this paper is organized as follows.
Section~\ref{sec:motivation} motivates the need to predict the full performance-cost trade-off of applications.
Section~\ref{sec:inference} describes the workflow of our prediction tool when making a performance prediction for a submitted application.
Section~\ref{sec:training} describes how our tool is deployed in a specific system, including how the training data is collected and how the fingerprint configurations are selected.
Section~\ref{sec:methodology} summarizes our methodology and Section~\ref{sec:evaluation} thoroughly evaluates our tool and the various design decisions made.
Finally, Section~\ref{sec:related} reviews related work and Section~\ref{sec:conclusion} concludes.

\section{Motivation}\label{sec:motivation}

\begin{figure}[h]
    \centering
    \includegraphics[width=\columnwidth]{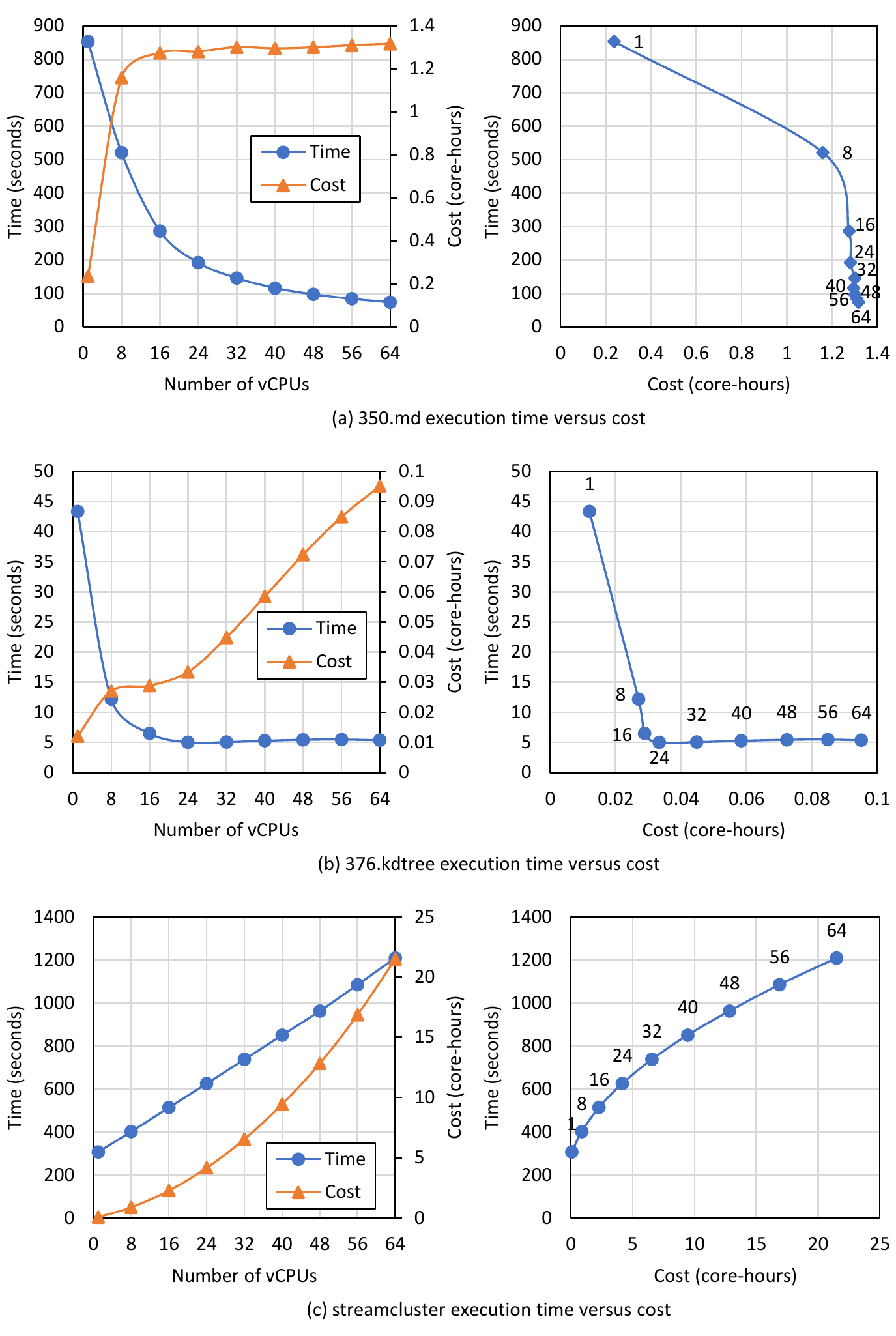}
    \caption{Trade-off between execution time and cost of three applications}\label{fig:motivation}
\end{figure}

Finding the best resource configuration for running an application is a challenging task and depends on the behavior of the application as well as the performance and cost objectives of the user.
Figure~\ref{fig:motivation} shows the trade-off between execution time and cost for three different applications (see Section~\ref{sec:methodology} for methodology).
We observe that each application exhibits a unique behavior that prompts different users to take different actions.

The first application, 350.md in Figure~\ref{fig:motivation}(a), witnesses a modest decrease in execution time when going from 1~vCPU to 8~vCPUs resulting in a substantial increase in cost.
However, beyond that point, the application begins to scale well experiencing a substantial improvement in performance with little cost increase.
For such an application, a performance-oriented user who is willing to reach the budget for 8~vCPUs might as well request the full 64~vCPUs since they would get substantial performance increase for little additional cost.

The second application, 376.kdtree in Figure~\ref{fig:motivation}(b), witnesses a substantial decrease in execution time when going from 1~vCPU to 16~vCPUs resulting in a minor increase in cost.
However, beyond that point, the execution time decreases negligibly resulting in a substantial increase in cost.
For such an application, even a performance-oriented user should not request more than 16~vCPUs since they would just be incurring additional cost for no performance benefit.

The third application, streamcluster in Figure~\ref{fig:motivation}(c), witnesses an increase in execution time when given more vCPUs.
For such an application the scales poorly, both a budget-oriented and a performance-oriented user should use only 1~vCPU to execute the application.

In each of the aforementioned cases, users need to be able to see the full trade-off space between execution time and performance to make the right decisions that meet their objectives.
Simply knowing the least-cost or the best-performing configuration is not sufficient for making nuance decisions.
This problem is further complicated when users have multiple systems available to them such that different Pareto-optimal choices may come from different systems.
For this reason, we propose a performance prediction tool that can predict the performance of an arbitrary application on multiple systems and configurations per system.

\section{Prediction Tool Workflow}\label{sec:inference}

In this section, we describe the workflow of our prediction tool after it has been deployed.
In other words, we describe how our tool makes a performance prediction for a submitted application.
We outline the workflow of our global and trade-off predictor in Section~\ref{sec:inference-flow}, then discuss the individual steps of that workflow in Sections~\ref{sec:inference-fingerprint}, \ref{sec:inference-classification}, and~\ref{sec:inference-regression}.
We discuss how we predict the application's sensitivity to interference in Section~\ref{sec:inference-interference}.
Finally, we describe how we trade-off the scope of prediction for higher prediction accuracy and lower online profiling overhead in Section~\ref{sec:local-tradeoff}.

\subsection{Global Trade-off Predictor}\label{sec:inference-flow}

\begin{figure}[ht]
    \centering
    \includegraphics[width=\columnwidth]{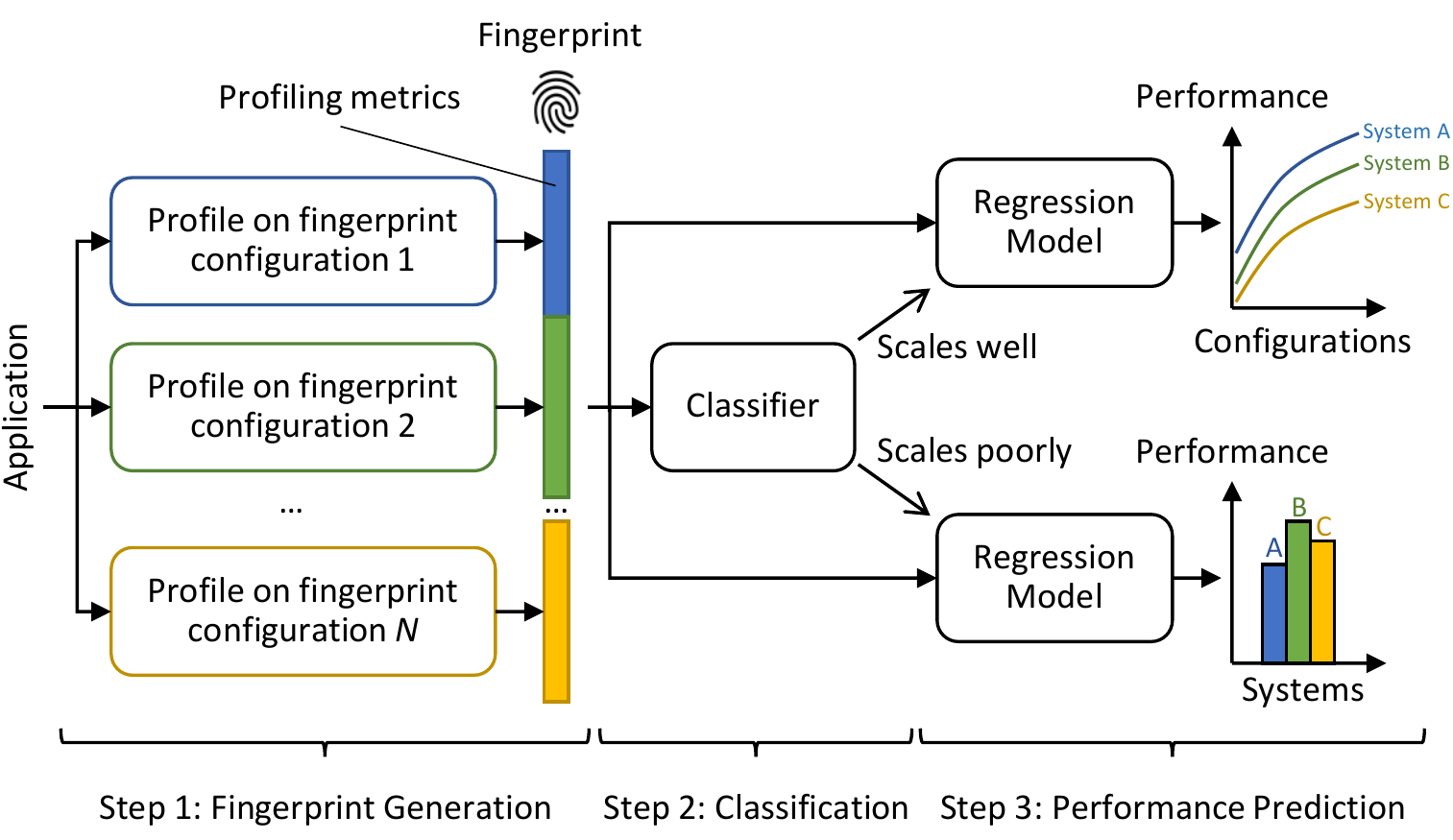}
    \caption{Global trade-off predictor workflow}
    \label{fig:flow}
\end{figure}

The global trade-off predictor predicts the performance-cost trade-off of an application across all configurations in all available systems.
Figure~\ref{fig:flow} shows the overall workflow of our global trade-off predictor.
In the first step, an application submitted to the tool is executed on a select set of configurations from the different systems, called \textit{fingerprint configurations}, to generate a fingerprint for the application.
An application's \textit{fingerprint} is a set of profiling metric values collected while executing the application on each of the fingerprint configurations.
The fingerprint is used as an input feature vector to the tool's prediction models.

In the second step, the fingerprint is passed to a classifier which classifies the application into one of two categories: scales well or scales poorly.
Depending on the result of the classification, the application will be treated differently in the third step.

In the third step, the fingerprint of the application is passed to one of two regression models depending on whether the application scales well or scales poorly.
For applications that scale well, the regression model predicts the performance of the application on all systems and configurations for some performance metric.
The performance metric used in the main evaluation in this paper is speedup relative to a baseline system and configuration.
This relative speedup can then be used to derive the relative cost and visualize the complete performance-cost trade-off space.
It can also be used to derive the absolute execution time and the absolute cost if the application is run to completion on one of the configurations.
For applications that scale poorly, it is unnecessary to make a prediction for all configurations on each system because it is always best to execute the application using the configuration with the fewest resources.
For this reason, the fingerprint is passed to a different regression model that predicts the relative performance of the application on all systems only for the configuration with the fewest resources.

\subsection{Fingerprint Generation}\label{sec:inference-fingerprint}

In the fingerprint generation step, the submitted application is executed and profiled on a set of fingerprint configurations to generate the application's fingerprint.
Important aspects of performing this step include: the choice of fingerprint configurations to execute on, the choice of profiling metrics to collect, and the span of the data collection.

\subsubsection{Choice of Fingerprint Configurations}
The number and choice of fingerprint configurations used for profiling the application depends on the set of systems being targeted and is decided at deployment time while the prediction models are being trained.
Using more fingerprint configurations has the advantage of improving prediction accuracy but with marginal returns as the number increases.
On the other hand, it has the disadvantage of requiring more execution resources to generate the fingerprint of a submitted application.
We describe how we select the fingerprint configurations at deployment time in Section~\ref{sec:training-global}.
We evaluate the impact of selecting the fingerprint configurations in Section~\ref{sec:evaluation-fingerprint}.

\subsubsection{Choice of Profiling Metrics}
The choice of profiling metrics to be collected depends on the system on which the profiling takes place.
Different systems may have different profiling tools and metrics available, so our tool supports using different metrics from different systems and configurations in the fingerprint.
In all cases, we use relative metrics (i.e.,~events per second) because they are independent of the application's absolute execution time.
Using metrics that are independent of absolute execution time has two advantages.
This first advantage is that it makes it possible to collect the metrics without running the application to completion.
The second advantage is that it makes it possible to train the models with a variety of applications having a wide range of execution times.
The specific set of metrics used from each system and configuration are decided at deployment time while the prediction models are being trained as part of the feature selection process.

\subsubsection{Span of Data Collection}
When running a submitted application on the fingerprint configurations, the application may be run partially or it may be run to completion.
Running the application to completion may be acceptable for short-running applications that execute frequently with the same workload size, such that the overhead of running to completion can be amortized.
In this case, the absolute and relative execution time on the fingerprint configurations can be measured and included in the fingerprint, which makes it easier for the models to make predictions for other configurations.
However, for long-running applications that do not execute frequently, running them to completion on all fingerprint configurations would require a large amount of execution resources and the cost is not amortized.
For this reason, it is desirable to be able to make predictions based on relative profiling metrics only without having to measure the application's absolute and relative execution time.
In the main evaluation in this paper, we use relative profiling metrics collected from partial runs because we prioritize not needing to run a submitted application to completion in order to make a prediction for it.
However, in Section~\ref{sec:evaluation-include-speedup}, we evaluate the impact of running the application to completion and including its relative execution time on the fingerprint configurations in the fingerprint.

\subsection{Classification by Scalability}\label{sec:inference-classification}

In the classification step, we classify submitted applications into those that scale well and those that scale poorly.
We define an application as scaling poorly if when scaled from the configuration with fewest resources to the configuration with the most resources, it slows down on the majority of systems (in practice, this behavior tends to be consistent across all systems).
If an application scales poorly, then giving this application more resources will cause it to incur substantially higher cost without witnessing any significant performance benefit.
An example of such an application is shown in Figure~\ref{fig:motivation}(c).
In this case, it is always best to execute this application on the configuration with the fewest resources, so it is not useful to make accurate predictions for other configurations.
It is only useful to predict the relative performance of this application across the different systems.
By separating the applications that scale well from those that scale poorly and using different regression models for each category, we allow each regression model to focus on having better accuracy for its category.

We use a random forest classifier to classify applications by scalability.
We evaluate the accuracy of the classifier in Section~\ref{sec:evaluation-classifier}.
This classifier is used in the main evaluation in this paper, however, Section~\ref{sec:evaluation-no-classifier} evaluates the impact of not using this classifier and using the same regression model for all applications.

\subsection{Performance Prediction}\label{sec:inference-regression}

In the performance prediction step, we pass the fingerprint of a submitted application to the appropriate regression model, and the regression model predicts the performance of the application on the different systems and configurations.
We use speedup over a baseline system and configuration as the predicted performance metric, but the model may also be trained to predict other metrics.
The baseline system and configuration are chosen during the fingerprint configuration selection process.
We use XGBoost~\cite{chen2016xgboost} for our regression models.
We evaluate the accuracy of our regression models in Section~\ref{sec:evaluation-regression}.

\subsection{Predicting Sensitivity to Interference}\label{sec:inference-interference}

Among the systems a user may have access to, some systems may be shared by other users who can run their applications simultaneously on the same node, such as in cloud environments.
For this reason, it is useful for users to know the extent to which the predicted performance may fluctuate in the presence of interference from other users.
To predict the sensitivity of applications to interference, we train another interference-aware regression model that predicts the performance of submitted applications on all systems and configurations as well as under different interference extremes for each system and configuration.
We make predictions for four types of interference patterns: no interference, compute-intensive interference, cache-intensive interference, and memory-intensive interference.
Hence, the model will produce four outputs for every system-configuration combination, one for each interference pattern.

The predictions made by this interference-aware regression model are useful because they provide users with a bound on the performance fluctuation they may observe if they run their application alongside other applications.
These predictions can also be useful to schedulers when making decisions on which applications to co-locate together.
Finally, even in non-shared environments, the same approach can be useful for predicting the sensitivity of an application to different QoS constraints.
For example, knowing how an application's performance may fluctuate when varying different QoS constraints can be useful for service providers in deciding what part of the system to throttle to meet a fluctuating power budget.
We evaluate the accuracy of our interference-aware regression model in Section~\ref{sec:evaluation-interference}.

\subsection{Scope of Prediction}\label{sec:local-tradeoff}

The global trade-off predictor described so far profiles an application on a specific set of systems and configurations to make a prediction for all systems and configurations.
However, in some cases, a user may be interested in a specific system, or even a specific range of configurations in one system.
In such cases, focusing on a narrower scope of prediction can allow for potentially higher accuracy and lower online profiling overhead.
For this reason, in addition to our global trade-off predictor, we also provide a single-system predictor and a local trade-off predictor.

The single-system predictor resembles the global trade-off predictor, except that it trains separate models for each system.
It also typically needs to run the application on fewer fingerprint configurations that come from just the system of interest.

\begin{figure}[ht]
    \centering
    \includegraphics[width=\columnwidth]{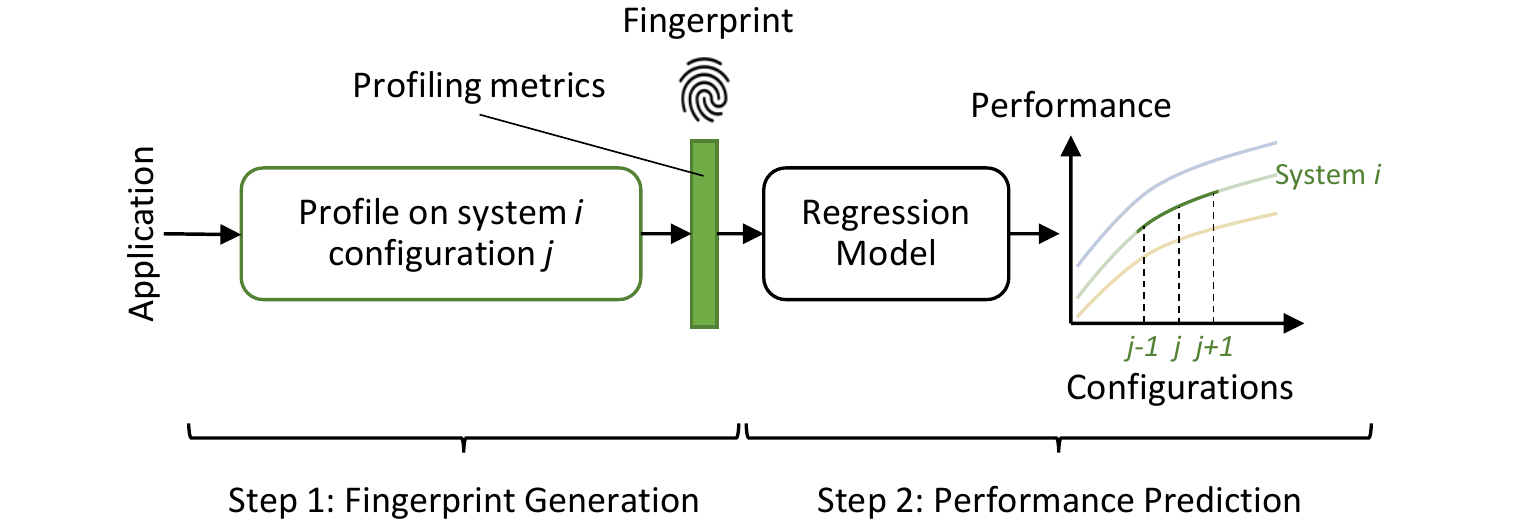}
    \caption{Local trade-off predictor workflow}
    \label{fig:flow-local}
\end{figure}

On the other hand, the local trade-off predictor profiles an application on a single configuration from a single system and predicts the relative performance of the application on nearby configurations.
Figure~\ref{fig:flow-local} shows the workflow of our local trade-off predictor.
In the first step, a submitted application is executed and profiled on a system and configuration specified by the user to generate a fingerprint consisting of the profiling metric values collected.
In the second step, the fingerprint is passed to a regression model specific to that system and configuration which predicts the relative performance of the application on the configurations with slightly more or slightly fewer cores.
Again, we use XGBoost~\cite{chen2016xgboost} as the regression model.

The motivation for providing this alternative predictor is that a user may already have an idea about which system and range of configurations are most suitable and only needs to fine tune that decision.
In this case, the local trade-off predictor can provide more accurate predictions and with less overhead since the application only needs to be profiled once.
The fingerprint could also be obtained from historical runs of the application.
In this case, our local predictor would be useful at making retrospective suggestions on the expected performance-cost trade-off if the user chooses to scale up or scale down in the future.

\section{Prediction Tool Deployment}\label{sec:training}

In this section, we describe the steps for deploying our prediction tool on a target set of systems.
We first describe how we collect the training data in Section~\ref{sec:training-data}.
We then describe how we train the prediction models in Section~\ref{sec:training-global}.

\subsection{Collection of Training Data}\label{sec:training-data}

The first step in deployment is to collect the data needed to train the models.
To do so, we execute a large number of benchmarks on the different systems and configurations and profile them while they execute.
The benchmarks are executed to completion and their execution time is measured so that their relative speedup on each system and configuration can be derived.
In the main evaluation in this paper, we execute each benchmark on all systems and configurations per system when collecting training data.
This approach has the advantage of giving a large amount of information to our prediction models so that they may have high accuracy.
However, it also requires a large number of executions to collect the training data.
Alternatively, each benchmark could be profiled on a random subset of the systems and configurations and the prediction models could be trained without the training data having exhaustive coverage.
We evaluate the impact of using non-exhaustive training data in Section~\ref{sec:evaluation-sparse}.

\subsection{Training the Global and Single-System Trade-off Models}\label{sec:training-global}

For the local trade-off predictor, we train a separate prediction model for each system and configuration.
The input to that prediction model is the set of profiling metrics collected from that system and configuration, and the output is the relative performance on the nearby configurations on the same system.

For the global and single-system trade-off predictors, one important aspect is selecting the number and choice of fingerprint configurations to use for generating the inputs to these models.
Trying all possible combinations of fingerprint configurations to find the optimal one would be prohibitively expensive.
For this reason, we use a greedy approach instead which works as follows.
We start by assuming that we will only use one fingerprint configuration.
We try all configurations on all systems as that one fingerprint configuration and select the one that results in the lowest error for the regression model for applications that scale well, fixing that as the first training configuration.
We then assume that we will use two fingerprint configurations.
We try all remaining configurations as the second fingerprint configuration and again select the one that results in the lowest error, fixing that as the second fingerprint configuration.
We keep repeating this process of adding more fingerprint configurations until the improvement in error drops below a threshold indicating that there is little marginal return from adding more fingerprint configurations.
The baseline configuration is selected in a similar manner.

\begin{table*}[t]
    \centering
    \caption{Metrics collected from each system and selected by the feature selection\\(numbers in parentheses indicate for which fingerprint configurations of the global predictor the metric was selected)}\label{tab:metrics}
    \resizebox{\textwidth}{!}{
        
\begin{tabular}{|c|c|c|c|c|c|c|c|c|c|c|}
\cline{1-3}\cline{5-7}\cline{9-11}
\textbf{System}&\textbf{Metrics}&\textbf{Selected}&&\textbf{System}&\textbf{Metrics}&\textbf{Selected}&&\textbf{System}&\textbf{Metrics}&\textbf{Selected}\\
\cline{1-3}\cline{5-7}\cline{9-11}
System 1&alignment-faults&No&&System 1&l2\_cache\_req\_stat.ls\_rd\_blk\_c&Yes (1)&&Systems 2 \& 3&fp\_arith\_inst\_retired.128b\_packed\_double&Yes (2, 3)\\
&all\_dc\_accesses&No&&(continued)&l2\_cache\_req\_stat.ls\_rd\_blk\_cs&No&&(continued)&fp\_arith\_inst\_retired.128b\_packed\_single&Yes (2)\\
&all\_tlbs\_flushed&No&&&l2\_dtlb\_misses&Yes (1)&&&fp\_arith\_inst\_retired.256b\_packed\_double&Yes (3)\\
&bp\_l1\_tlb\_miss\_l2\_hit&Yes (1)&&&l2\_itlb\_misses&Yes (1)&&&fp\_arith\_inst\_retired.256b\_packed\_single&Yes (2)\\
&bp\_l1\_tlb\_miss\_l2\_tlb\_miss&Yes (1)&&&l2\_latency.l2\_cycles\_waiting\_on\_fills&No&&&fp\_arith\_inst\_retired.512b\_packed\_double&Yes (3)\\
&branch-instructions&Yes (1)&&&l2\_request\_g1.cacheable\_ic\_read&Yes (1)&&&fp\_arith\_inst\_retired.512b\_packed\_single&Yes (3)\\
&branch-misses of all branches&No&&&l2\_request\_g1.ls\_rd\_blk\_c\_s&Yes (1)&&&fp\_arith\_inst\_retired.scalar\_double&No\\
&cache-misses of all cache refs&Yes (1)&&&l2\_request\_g1.rd\_blk\_l&Yes (1)&&&fp\_arith\_inst\_retired.scalar\_single&Yes (2, 3)\\
&cache-references&Yes (1)&&&l2\_request\_g1.rd\_blk\_x&No&&&fp\_assist.any&Yes (3)\\
&context-switches&No&&&l2\_request\_g2.bus\_locks\_originator&No&&&GHz&Yes (2, 3)\\
&cpu-clock CPUs utilized&Yes (1)&&&ls\_l1\_d\_tlb\_miss.tlb\_reload\_1g\_l2\_hit&No&&&insn per cycle&Yes (3)\\
&cpu-cycles GHz&No&&&ls\_tlb\_flush&No&&&inst\_retired.any&No\\
&cpu-migrations&No&&&major-faults&Yes (1)&&&iTLB-load-misses&No\\
&de\_dis\_dispatch\_token\_stalls0.alu\_token\_stall&Yes (1)&&&minor-faults&Yes (1)&&&iTLB-loads&Yes (2, 3)\\
&de\_dis\_dispatch\_token\_stalls0.retire\_token\_stall&Yes (1)&&&page-faults&No&&&L1-dcache-load-misses&Yes (3)\\
&de\_dis\_dispatch\_token\_stalls1.load\_queue\_token\_stall&No&&&sse\_avx\_stalls&No&&&L1-dcache-loads&Yes (2, 3)\\
&de\_dis\_dispatch\_token\_stalls1.store\_queue\_token\_stall&Yes (1)&&&stalled cycles per insn&Yes (1)&&&L1-dcache-store&No\\
&dTLB-load-misses of all dTLB cache accesses&Yes (1)&&&stalled-cycles-backend backend cycles idle&No&&&l2\_rqsts.all\_pf&Yes (2, 3)\\
&dTLB-loads&No&&&stalled-cycles-frontend frontend cycles idle&Yes (1)&&&l2\_rqsts.all\_rfo&Yes (3)\\
&ex\_ret\_instr&Yes (1)&&&task-clock CPUs utilized&Yes (1)&&&l2\_rqsts.miss&Yes (3)\\
&ex\_ret\_mmx\_fp\_instr.mmx\_instr&No&&&uops\_dispatched&Yes (1)&&&l2\_rqsts.references&Yes (3)\\
&ex\_ret\_mmx\_fp\_instr.sse\_instr&No&&&uops\_retired&No&&&l2\_rqsts.rfo\_hit&Yes (2, 3)\\
\cline{5-7}
&fp\_ret\_sse\_avx\_ops.all&No&&Systems 2 \& 3&branches&No&&&l2\_rqsts.rfo\_miss&Yes (3)\\
&ic\_fetch\_stall.ic\_stall\_any&Yes (1)&&&branch-misses&Yes (2)&&&LLC-load-misses&Yes (2, 3)\\
&ic\_fetch\_stall.ic\_stall\_back\_pressure&Yes (1)&&&bus-cycles&Yes (2)&&&LLC-loads&Yes (2, 3)\\
&ic\_fetch\_stall.ic\_stall\_dq\_empty&Yes (1)&&&cache-misses&Yes (2, 3)&&&LLC-store-misses&No\\
&instructions insn per cycle&No&&&cache-references&Yes (3)&&&LLC-stores&Yes (3)\\
&iTLB-load-misses of all iTLB cache accesses&No&&&context-switches&No&&&longest\_lat\_cache.miss&Yes (2, 3)\\
&iTLB-loads&Yes (1)&&&cpu-migrations&Yes (2, 3)&&&longest\_lat\_cache.reference&Yes (2)\\
&l1\_dtlb\_misses&Yes (1)&&&CPUs utilized&Yes (2, 3)&&&mem\_inst\_retired.all\_loads&Yes (3)\\
&L1-dcache-load-misses of all L1-dcache accesses&Yes (1)&&&cycle\_activity.cycles\_l1d\_miss&Yes (3)&&&mem\_inst\_retired.all\_stores&Yes (2)\\
&L1-dcache-loads&No&&&cycle\_activity.cycles\_l2\_miss&Yes (2, 3)&&&mem\_load\_retired.l1\_hit&Yes (2, 3)\\
&L1-dcache-prefetches&Yes (1)&&&cycle\_activity.cycles\_l3\_miss&Yes (2)&&&mem\_load\_retired.l1\_miss&Yes (2)\\
&L1-icache-load-misses of all L1-icache accesses&Yes (1)&&&cycle\_activity.cycles\_mem\_any&Yes (2)&&&mem\_load\_retired.l2\_hit&Yes (2)\\
&L1-icache-loads&No&&&cycle\_activity.stalls\_l1d\_miss&No&&&mem\_load\_retired.l2\_miss&Yes (2)\\
&l2\_cache\_accesses\_from\_dc\_misses&No&&&cycle\_activity.stalls\_l2\_miss&Yes (2)&&&mem\_load\_retired.l3\_hit&Yes (2)\\
&l2\_cache\_accesses\_from\_ic\_misses&Yes (1)&&&cycle\_activity.stalls\_l3\_miss&Yes (3)&&&mem\_load\_retired.l3\_miss&Yes (2, 3)\\
&l2\_cache\_hits\_from\_dc\_misses&No&&&cycle\_activity.stalls\_mem\_any&No&&&mem-loads&Yes (3)\\
&l2\_cache\_hits\_from\_ic\_misses&No&&&dTLB-load-misses&Yes (3)&&&mem-stores&Yes (2, 3)\\
&l2\_cache\_hits\_from\_l2\_hwpf&No&&&dTLB-loads&Yes (3)&&&page-faults&Yes (2, 3)\\
&l2\_cache\_misses\_from\_dc\_misses&Yes (1)&&&dTLB-store-misses&Yes (2, 3)&&&resource\_stalls.any&Yes (3)\\
&l2\_cache\_misses\_from\_ic\_miss&No&&&dTLB-stores&No&&&&\\
\cline{1-3}\cline{5-7}\cline{9-11}
\end{tabular}

    }
\end{table*}

By \textit{trying} a set of fingerprint configurations, we mean generating the fingerprints using those configurations and training the prediction models using those fingerprints.
Throughout this process, we use all the collected profiling metrics as part of the fingerprint because trying to jointly select fingerprint configurations and profiling metrics would be prohibitively expensive.
Once we have finalized the set of fingerprint configurations, we apply standard feature selection techniques to reduce the number of profiling metrics needed from each configuration.
A different number and set of metrics may be selected from each configuration.
Feature selection improves prediction accuracy and also reduces the number of metrics that need to be collected online, thereby reducing the online profiling overhead.

\section{Methodology}\label{sec:methodology}

We evaluate our tool using three single-node CPU systems with the following specifications:
\begin{itemize}
    \item System 1: A dual socket AMD EPYC 7532 CPU system with 64 cores (64 threads) and 512GB of main memory
    \item System 2: A dual socket Intel Xeon Gold 6242 CPU system with 32 cores (64 threads) and 192GB of main memory
    \item System 3: A dual socket Intel Xeon Gold 5120 CPU system with 28 cores (56 threads) and 64GB of main memory
\end{itemize}
On each system, the configurations considered are 1~vCPU and all multiple of 8~vCPUs, with the main memory distributed evenly across~vCPUs.
For example, System 1 has the following configurations: 1~vCPU~+~8~GB, 8~vCPUs~+~64~GB, 16~vCPUs~+~128~GB, 24~vCPUs~+~192~GB, 32~vCPUs~+~256~GB, 40~vCPUs~+~320~GB, 48~vCPUs~+~384~GB, 56~vCPUs~+~448~GB, and 64~vCPUs~+~512~GB.
Overall, the three systems together have a total of 26 configurations.

In the evaluation in this paper, we use \texttt{docker}~\cite{docker} to configure the number of vCPUs and the amount of physical memory to be given to each application.
We also use \texttt{perf}~\cite{perf} to profile the benchmarks and collect a large number of profiling metrics.
However, our workflow and prediction models are not specific to \texttt{docker} and \texttt{perf}.
Other means of configuring resources and collecting profiling metrics may also be used, and our training and prediction workflows would operate in the same way.

We collect around 60 profiling metrics from each system when obtaining the training data.
These metrics differ across systems because different CPUs have different counters available.
The metrics collected from each system are shown in Table~\ref{tab:metrics}.
The table also shows which of these metrics were kept after feature selection for each of the fingerprint configurations used in the global trade-off predictor.
We measured the performance overhead of collecting 60 profiling metrics using \texttt{perf} to be less than 2\% on average.

For predicting the sensitivity of applications to interference, we use \texttt{stress-ng}~\cite{stress-ng} to simulate compute-intensive, cache-intensive, or memory-intensive interference on the unused cores in the system.
However, our approach is not specific to \texttt{stress-ng} and other tools for simulating interference (or QoS constraints) can also be used without affecting our workflow.

To train and test our prediction models, we use 69 data analytics and scientific computing benchmarks from seven benchmarks suites and libraries.
These benchmarks are listed in Table~\ref{tab:benchmarks}.

\begin{table}[ht]
    \centering
    \caption{Benchmarks used in the evaluation}\label{tab:benchmarks}
    \resizebox{\columnwidth}{!}{
        
\begin{tabular}{|p{1.1in}|p{1.8in}|}
    \hline
    \textbf{Suite} & \textbf{Benchmarks} \\
    \hline
    NAS Parallel Benchmarks~\cite{bailey1991parallel} & bt, cg, ep, ft, is, lu, mg, sp \\ \hline
    PARSEC3.0~\cite{zhan2017parsec3} & blackscholes, bodytrack, canneal, freqmine, streamcluster, swaptions, splash2.barnes, splash2.cholesky, splash2.radiosity, splash2.volrend    \\ \hline
    SPEC ACCEL~\cite{juckeland2014spec} & 550.pmd, 552.pep, 554.pcg, 555.pseismic, 556.psp, 559.pmniGhost, 560.pilbdc, 563.pswim, 570.pbt \\ \hline
    SPEC OMP 2012~\cite{aslot2001specomp} & 350.md, 351.bwaves, 352.nab, 357.bt, 358.botsalgn, 359.botsspar, 360.ilbdc, 363.swim, 367.imagick, 370.mgrid, 371.applu, 372.smithwa, 376.kdtree \\ \hline
    Parboil~\cite{stratton2012parboil} & bfs, cutcp, histo, lbm, mri-gridding, sgemm, spmv, stencil, tpacf \\ \hline
    Rodinia~\cite{che2009rodinia} & bfs, heartwall, hotspot, kmeans, leukocyte, lud\_omp, needle, pathfinder, srad\_v2 \\ \hline
    MLlib~\cite{meng2016mllib} & correlation, dtclassifier, fmclassifier, gbtclassifier, gmm, kmeans, logisticregression, lsvc, mlp, pca, randomforestclassifier, summarizer     \\ \hline
\end{tabular}

    }
\end{table}

Our training and inference workflows are implemented in Python.
We use NumPy~\cite{harris2020array} and pandas~\cite{reback2020pandas} for performing the data preparation and processing tasks.
We use the Scikit-Learn~\cite{pedregosa2011scikit} and XGBoost~\cite{chen2016xgboost} libraries for the classification, regression, and feature selection algorithms.

We use ten-fold cross-validation throughout the evaluation section to evaluate the prediction error of our models.
For each fold, we use fingerprints from full runs and partial runs for the benchmarks in the training set, and fingerprints from partial runs for the benchmarks in the testing set.
For the partial runs, the benchmarks are fingerprinted for a duration of 30~seconds in our experiments.
However, users may vary the fingerprinting duration to trade off accuracy for fingerprinting overhead.

We use the symmetric mean absolute percentage error (SMAPE)~\cite{smape} as the main error metric to guide our optimization.
It is possible to use other error metrics such as the mean absolute percentage error (MAPE), but we opted for SMAPE due its boundedness between 0\% and 200\% in addition to the symmetric penalty given whether the predictions are higher or lower than the true values (since we are predicting ratios).

\section{Evaluation}\label{sec:evaluation}

\subsection{Classification Accuracy}\label{sec:evaluation-classifier}

Table~\ref{tab:classifier-confusion} shows the confusion matrix for our random forest classifier which classifies benchmarks based on their scalability in the global trade-off predictor.
The results show that our classifier is highly accurate, correctly identifying 58 out of 60 applications that scale well, and eight out of nine applications that scale poorly.
Our tool is thus effective at screening out applications that scale poorly to allow the main regression model to focus on applications that scale well.

\begin{table}[ht]
    \centering
    \caption{Confusion matrix of the scalability classifier}\label{tab:classifier-confusion}
    
\begin{tabular}{cc|c|c|}
     & \multicolumn{1}{c}{} & \multicolumn{2}{c}{Predicted scalability} \\
     & \multicolumn{1}{c}{} & \multicolumn{1}{c}{Scales well} & \multicolumn{1}{c}{Scales poorly} \\
    \cline{3-4}
    True & Scales well & 58 & 2 \\
    \cline{3-4}
    scalability & Scales poorly & 1 & 8 \\
    \cline{3-4}
\end{tabular}

\end{table}

\subsection{Regression Model Accuracy}\label{sec:evaluation-fingerprint}\label{sec:evaluation-regression}

\begin{figure}[ht]
    \centering
    \includegraphics[width=\columnwidth]{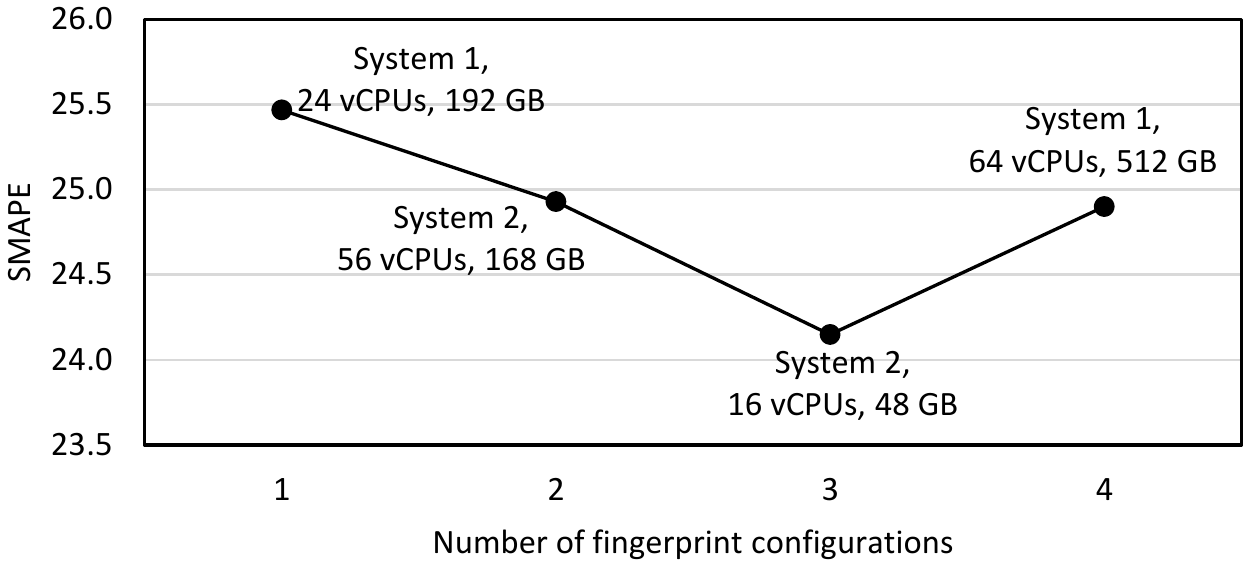}
    \caption{Change in regression error of the global trade-off predictor as the number of fingerprint configurations increases (labels indicate which configuration is selected each iteration)}\label{fig:evaluation-fingerprint}
\end{figure}

Figure~\ref{fig:evaluation-fingerprint} shows how the global trade-off predictor's regression error for applications that scale well varies as the number of fingerprint configurations is increased.
The points in the figure are labelled with the best fingerprint configuration found at each iteration.

Based on the results in Figure~\ref{fig:evaluation-fingerprint}, we make two key observations.
The first observation is that the regression model error decreases steadily as we go from using one fingerprint configuration to using three fingerprint configurations, reaching a mean error of 24.2\%.
However, increasing the number of fingerprint configurations beyond three hurts accuracy because the model becomes overloaded with information.
For this reason, we fix the number of fingerprint configurations to three (out of a total of 26).
Note that after feature selection is applied, the mean error is further reduced to 22.5\%.

The second observation is that the three selected fingerprint configurations span two different systems and three different vCPU counts.
This result shows that the fingerprint configuration selection process automatically diversifies the chosen set of fingerprint configurations.
Note, however, that one of the three systems does not appear in any of the fingerprint configurations.
This result shows that our tool can make predictions for a system without running the application on that system at all.

The aforementioned results show that our tool can achieve reasonable prediction accuracy across multiple systems by profiling an application on just three configurations.
However, in some cases, a user may already have a particular system in mind and is not interested in relative performance across other systems.
For such users, our tool also provides single-system regression models for each system that make performance-cost trade-off predictions for just that system.
Table~\ref{tab:single-system-fp} shows how the prediction errors of these single-system models vary as the number of fingerprint configurations increases.
We observe that by constraining the problem to a single system, our tool can achieve lower prediction error, even if profiling the application on a smaller number of configurations.
Note that after feature selection is applied, the mean error when using three fingerprint configurations is further reduced to 11.4\%, 12.5\%, and 15.6\% for System 1, 2, and 3, respectively.

\begin{table}[ht]
    \centering
    \caption{Change in regression error of single-system models as the number of fingerprint configurations increases}\label{tab:single-system-fp}
    \resizebox{\columnwidth}{!}{
        
\begin{tabular}{|l|l|c|c|c|c|}
    \hline
    \multicolumn{2}{|l|}{Number of configurations} & 1 & 2 & 3 & 4 \\
    \hline
    System 1 & Error & 13.5 & 13.2 & 12.3 & 12.3 \\
    \cline{2-6}
             & Configuration & 24 vCPUs & 32 vCPUs, & 8 vCPUs, & 56 vCPUs \\
             &               & 192GB    & 256GB     & 64GB     & 448GB    \\
    \hline
    System 2 & Error & 15.5 & 13.6 & 13.4 & 13.4 \\
    \cline{2-6}
             & Configuration & 16 vCPUs & 56 vCPUs & 40 vCPUs & 48 vCPUs  \\
             & & 48 GB & 168GB & 120GB & 144GB  \\
    \hline
    System 3 & Error & 19.0 & 17.4 & 17.0 & 17.1 \\
    \cline{2-6}
             & Configuration & 24 vCPUs & 32 vCPUs & 40 vCPUs & 8 vCPUs \\
             & & 24GB & 32GB & 40GB & 8GB \\
    \hline
\end{tabular}

    }
\end{table}

\begin{figure}[ht]
    \centering
    \includegraphics[width=\columnwidth]{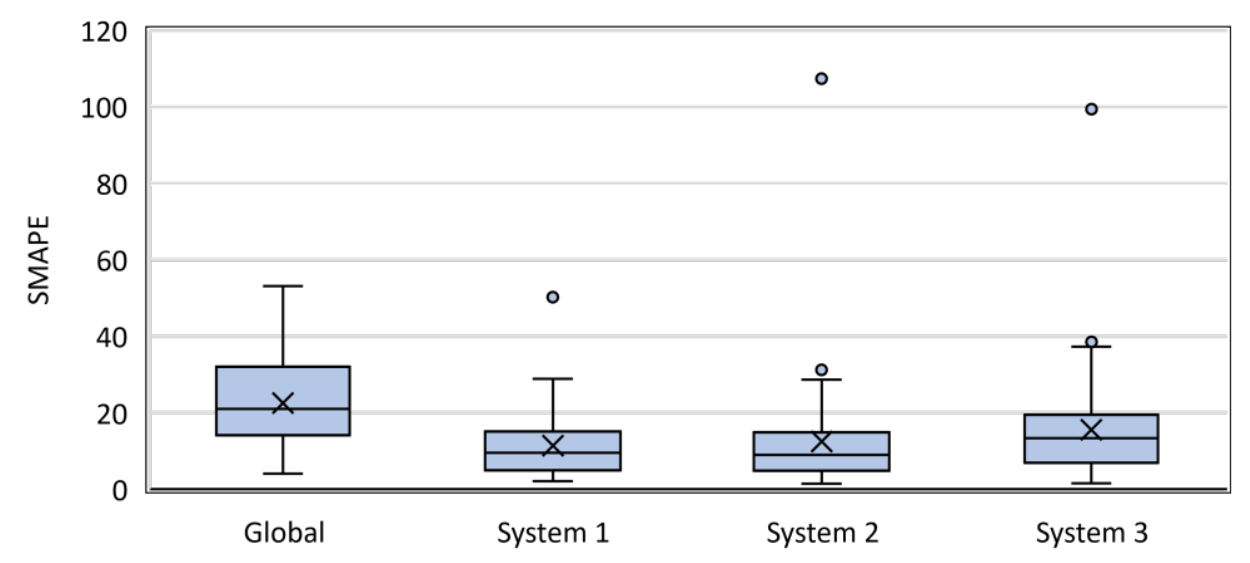}
    \caption{Distribution of error across benchmarks for the global and single-system predictors}\label{fig:error-distribution}
\end{figure}

The prediction error achieved by the global and single-system predictors varies across benchmarks.
Figure~\ref{fig:error-distribution} shows the distribution of the prediction error across benchmarks for each predictor.
It is clear that our models can make predictions for many benchmarks with very low error.
Moreover, the median error is consistently lower than the mean error, indicating that there are more benchmarks whose prediction error is below the mean.
However, the prediction error remains high for some benchmarks, especially outliers.
Screening out such benchmarks with more sophisticated classification is the subject of future work.

\subsection{Application Case Study: GROMACS}\label{sec:evaluation-case-studies}

\begin{figure}[ht]
    \centering
    \includegraphics[width=\columnwidth]{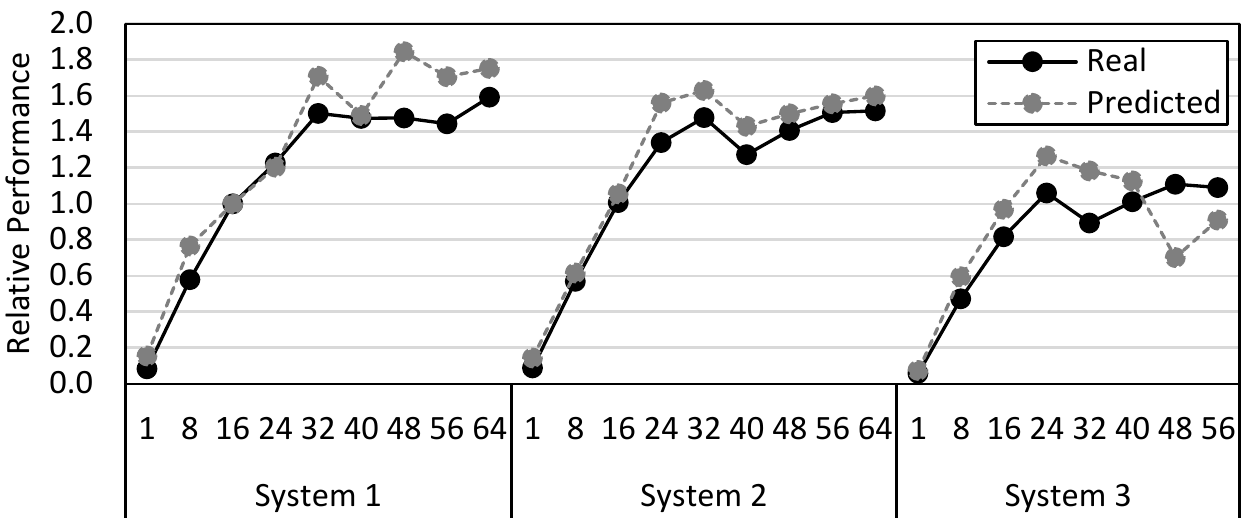}
    \caption{Real vs. predicted relative performance for GROMACS~\cite{van2005gromacs}}\label{fig:gromacs}
\end{figure}

To test the effectiveness of our tool in a realistic scenario, we apply it to a commonly used molecular dynamics simulation application called GROMACS~\cite{van2005gromacs}.
We use a global trade-off prediction model that has been trained using the benchmarks in Table~\ref{tab:benchmarks} and that has never seen GROMACS before.
We execute and profile GROMACS on the model's fingerprint configurations for only 5\% of its total execution time on the fastest system and configuration.
The fingerprint obtained is then passed to the model which predicts the relative performance of GROMACS on all the systems and configurations.
Figure~\ref{fig:gromacs} shows the real and predicted relative performance values.
It is clear that our tool does a good job at predicting how GROMACS behaves across the three systems.
Overall, our tool achieves a mean prediction error of 17.3\% for GROMACS.

\subsection{Predicting Sensitivity to Interference}\label{sec:evaluation-interference}

Recall from Section~\ref{sec:inference-interference} that our tool also provides interference-aware regression models that predict an application's performance on each system and configuration under different interference extremes to give users a sense of the application's sensitivity to different types of interference.
Table~\ref{tab:results-interference} shows the prediction error achieved for the global as well as the single-system predictors when predicting performance under each of the three considered interference extremes.
It is clear that the achieved prediction error is comparable to that achieved when no interference is assumed, albeit slightly higher.
It is noteworthy that for most systems, cache-intensive interference seems to be the most difficult type of interference to predict for, having a slightly higher error than the rest.

\begin{table}[ht]
    \centering
    \caption{Global and single-system regression error when predicting performance under different interference extremes}\label{tab:results-interference}
    \resizebox{\columnwidth}{!}{
        
\begin{tabular}{|c|c|c|c|}
    \hline
    Predictor & \multicolumn{3}{|c|}{Type of interference} \\
    \cline{2-4}
     & Compute-intensive & Memory-intensive & Cache-intensive \\
    \hline
    Global & 24.7 & 24.6 & 26.1 \\
    \hline
    System 1 only & 14.2 & 13.0 & 16.1 \\
    \hline
    System 2 only & 15.5 & 14.1 & 17.9 \\
    \hline
    System 3 only & 22.9 & 19.4 & 14.8 \\
    \hline
\end{tabular}

    }
\end{table}

\subsection{Impact of Having a Classification Stage}\label{sec:evaluation-no-classifier}

\begin{figure}[ht]
    \centering
    \includegraphics[width=\columnwidth]{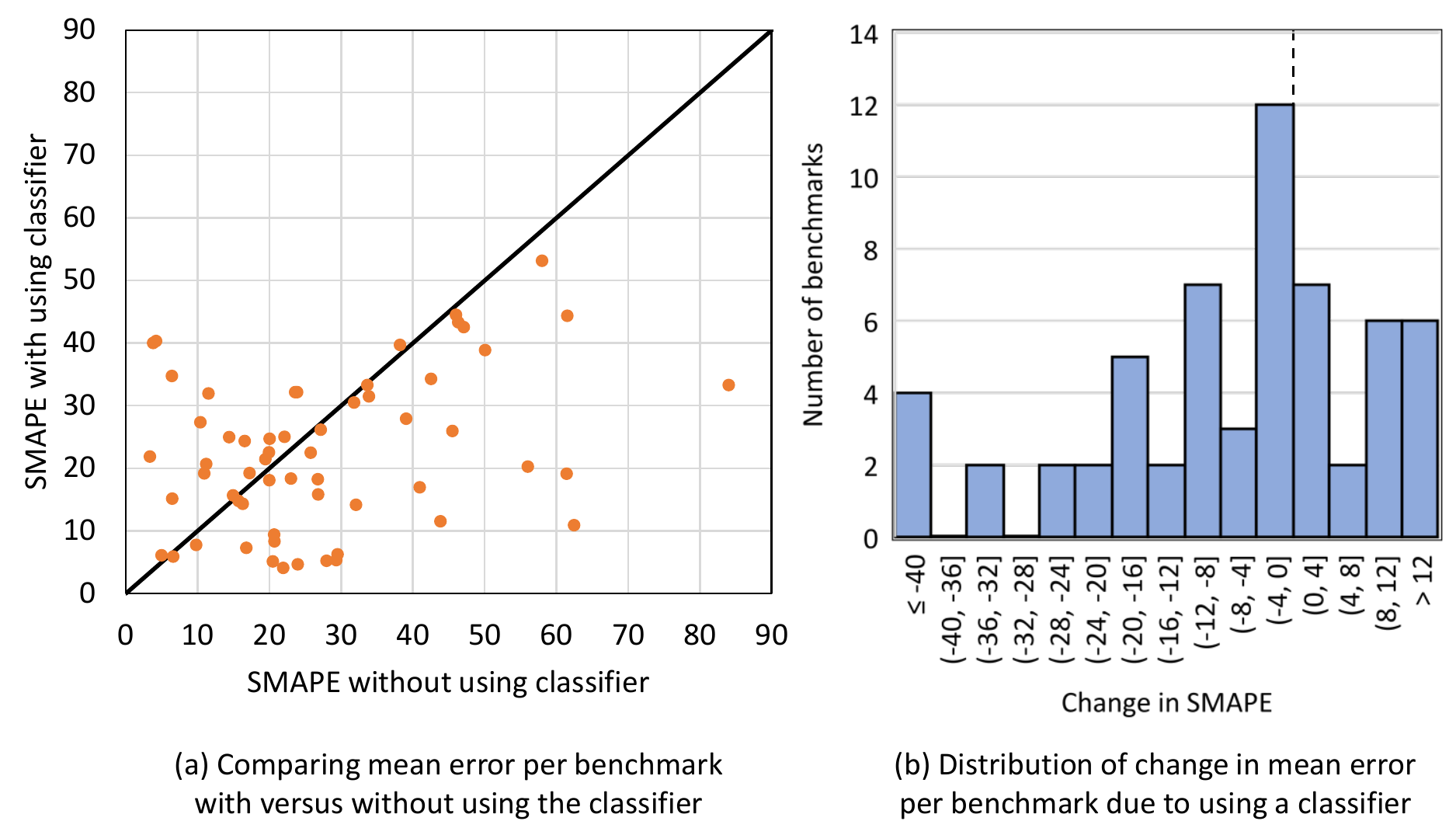}
    \caption{Impact of using a classifier before the regression model on the regression model error}\label{fig:no-classifier}
\end{figure}

Recall that one of the steps in our global trade-off predictor's workflow is using a classifier to distinguish applications that scale well from those that scale poorly to allow the main regression model to focus on applications that scale well.
In this subsection, we evaluate the benefit of this proposed classifier.
Figure~\ref{fig:no-classifier}(a) compares the mean regression model prediction error with and without using a classifier.
Each point in the figure represents a benchmark.
It is clear that the majority of benchmarks witness a reduction in mean error due to the use of the classifier.
Figure~\ref{fig:no-classifier}(b) shows the distribution of the change in mean error per benchmark due to the use of the classifier.
It is again clear that the majority of benchmarks witness a reduction in mean error.
The mean and median change in error across all benchmarks is -6.67\% and -2.25\%, respectively.
The overall mean error if a classifier were not used would be 29.2\%.
These results verify the effectiveness of our proposed classification stage at improving the quality of the regression model.

\subsection{Impact of Fingerprinting with Partial vs. Complete Runs}\label{sec:evaluation-include-speedup}

In our main evaluation, we only use relative profiling metrics collected from partial runs of the applications on the fingerprint configurations.
However, if an application is run to completion on the fingerprint configurations, its relative performance across the fingerprint configurations can be derived and included in the fingerprint, which gives more information to the regression model to make accurate predictions.
In this subsection, we evaluate the impact of using complete runs instead of partial runs on regression model accuracy.

\begin{figure}[ht]
    \centering
    \includegraphics[width=\columnwidth]{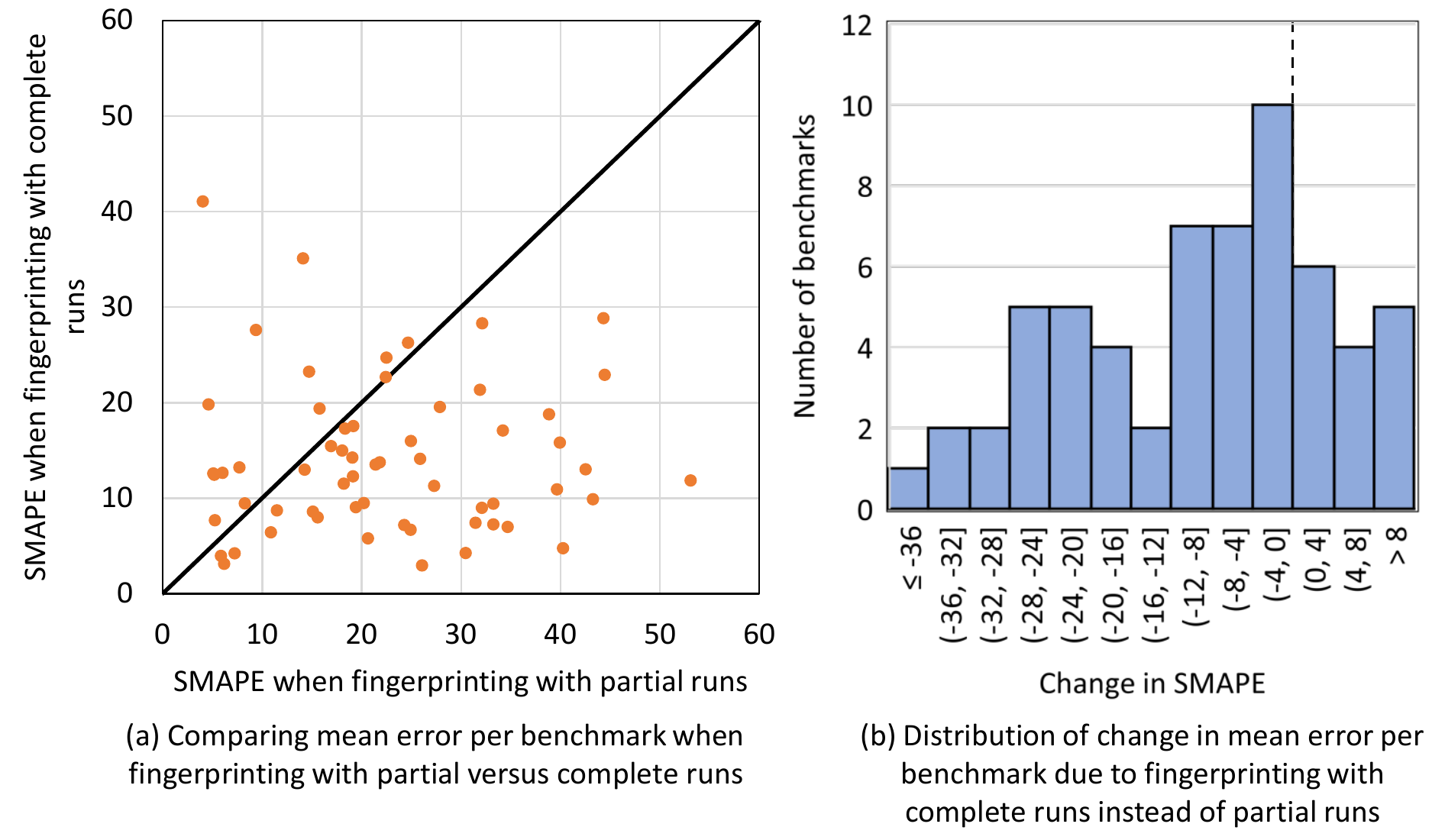}
    \caption{Impact of fingerprinting with partial runs versus complete runs}\label{fig:include-speedup}
\end{figure}

Figure~\ref{fig:include-speedup}(a) compares the mean regression model prediction error when fingerprinting with partial runs versus complete runs.
Each point in the figure represents a benchmark.
It is clear that the majority of benchmarks witness a reduction in mean error when complete runs are used.
Figure~\ref{fig:include-speedup}(b) shows the distribution of the change in mean error per benchmark due to the use of complete runs instead of partial runs when fingerprinting.
It is again clear that the majority of benchmarks witness a reduction in mean error.
The mean and median change in error across all benchmarks is -8.44\% and -7.27\%, respectively.
The overall mean error if performance was measured on the fingerprint configurations and included in the fingerprint would be 14.1\%.
These results demonstrate that a deployer of our tool can substantially reduce prediction error if they are willing to run submitted applications to completion when fingerprinting them.
However, using partial runs still provides reasonably accurate predictions.

\subsection{Impact of Training with Partial Data Coverage}\label{sec:evaluation-sparse}

In our main evaluation, we collect data to train the regression model by running every benchmark on every system and configuration.
This exhaustive coverage of training data demands a large amount of execution resources.
In this subsection, we evaluate the impact of using partial training data coverage by running each benchmark on a different random subset of all the systems and configurations.

\begin{figure}[b]
    \centering
    \includegraphics[width=\columnwidth]{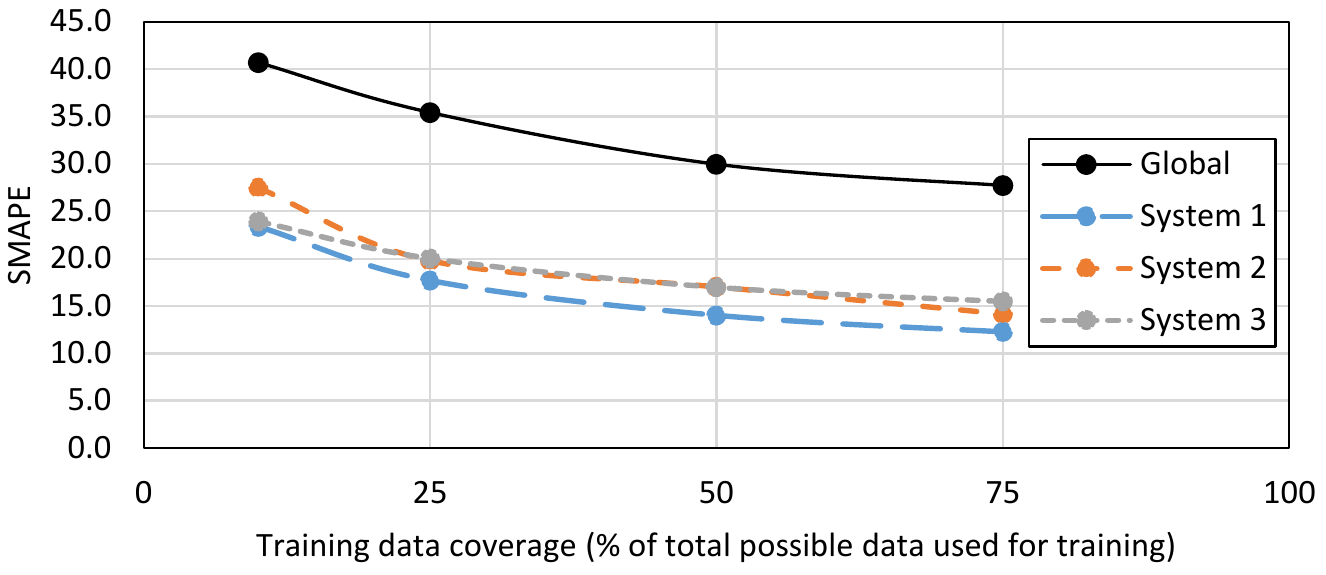}
    \caption{Impact of training with partial data coverage on global and single-system regression model accuracy}\label{fig:sparse-training}
\end{figure}

Figure~\ref{fig:sparse-training} shows how the regression model prediction accuracy varies with training data coverage for the global and single-system regression models.
As expected, the prediction error increases as the training data coverage decreases, but it does so more gradually for the single-system models than for the global model.
In fact, in the single-system models, the error remains under 20\% even with coverage as low as 25\%.
These result shows that deployers of our tool can trade-off accuracy for the amount of resources used during training data collection depending on the desired prediction quality.

The configurations used for each benchmark when training with partial data coverage were selected at random in this experiment.
One could further improve accuracy when training with partial data coverage by employing more sophisticated techniques for selecting which configuration to use for each benchmark during data collection.
Such techniques are the subject of our future work.

\subsection{Local Trade-off Predictor Accuracy}

Recall from Section~\ref{sec:evaluation-regression} that our global trade-off predictor achieves a prediction error of 22.5\% when predicting the trade-off space across three systems by profiling the application on just four fingerprint configurations.
Furthermore, by narrowing the scope of prediction, our single-system models achieve prediction errors between 11.4\% and 15.6\% while profiling the applications on just two fingerprint configurations.
In this subsection, we evaluate the local trade-off predictor described in Section~\ref{sec:local-tradeoff} which narrows the prediction scope even more in an attempt to achieve even lower prediction error while profiling the application on just a single configuration.

\begin{figure}[ht]
    \centering
    \includegraphics[width=\columnwidth]{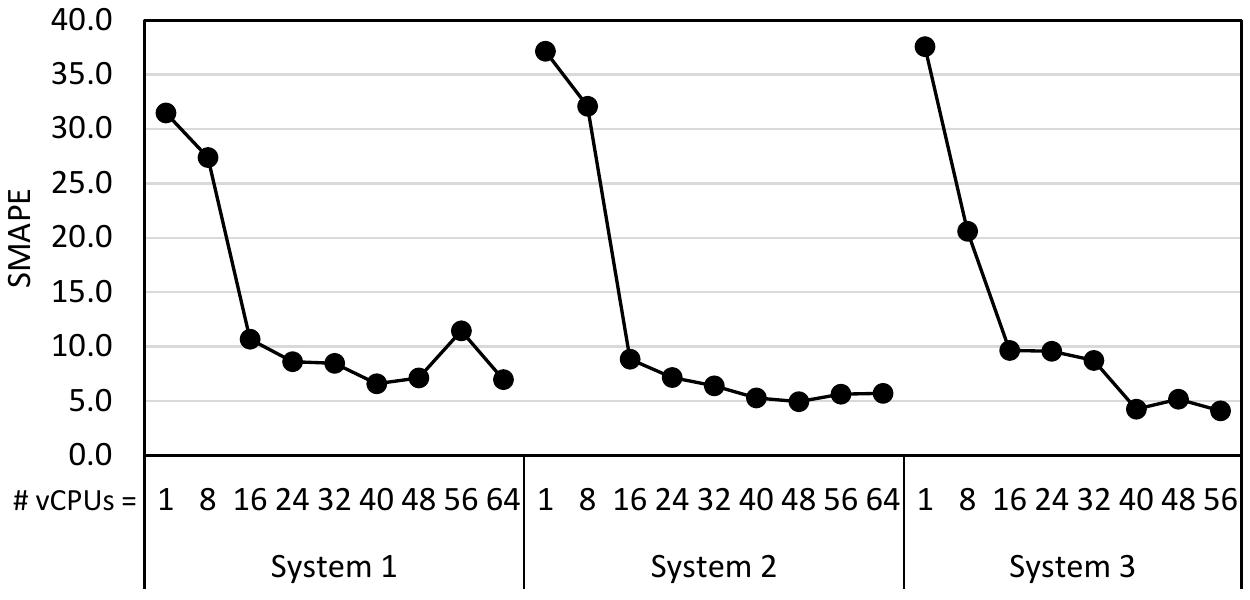}
    \caption{Prediction error of the local trade-off predictor for each configuration on each system}\label{fig:local-error}
\end{figure}

Figure~\ref{fig:local-error} shows the mean prediction error of each system and configuration's regression model in the local trade-off predictor.
Here, the error reported for each system and configuration's regression model is the error of predicting the relative performance of applications on the nearby configurations on the same system.
We observe that in the majority of cases, the models achieve a mean prediction error of under 10\%.
This result shows the effectiveness of the local trade-off predictor at reducing prediction error by narrowing the scope of prediction.

One interesting observation from Figure~\ref{fig:local-error} is that the prediction error for the configurations with 1~vCPU and 8~vCPUs is consistently high.
In particular, the predictor finds it difficult to predict the relative performance between the configurations with 1~vCPU and 8~vCPUs.
Upon investigating the benchmarks, we identified two reasons that make the transition between 1~vCPU and 8~vCPUs have an especially unpredictable performance impact.
On the one hand, when applications run with 1~vCPU, they do not incur any parallelization overhead which gives this configuration a performance advantage compared to the remaining configurations.
On the other hand, when applications run with 1~vCPU, they have a tighter memory budget which gives this configuration a performance disadvantage.

\section{Related Work}\label{sec:related}

There is a large body of literature on using prediction tools to assist users with selecting the best system and/or configuration to execute their applications.
One major direction is to find the optimal or near-optimal system and configuration that minimize some objective function~\cite{cherrypick,hsu2018arrow,liu2019accordia,klimovic2018selecta,hsu2018micky,xiao2019sara,vanir,casimiro2020lynceus,bilal2020best,hsu2018scout,lin2020fife,wu2021apollo} by using optimization techniques.
For example, CherryPick~\cite{cherrypick} uses Bayesian optimization to find near-optimal configurations for big data analytics applications in the public cloud.
Arrow~\cite{hsu2018arrow} augments Bayesian optimization with low-level performance information to reduce its fragility.
Accordia~\cite{liu2019accordia} improves the performance of Gaussian-process-based approaches by using the upper confidence bound technique and by supporting an early abort mechanism for non-promising configurations.
Selecta~\cite{klimovic2018selecta} finds near-optimal configurations for both compute and storage together.
Micky~\cite{hsu2018micky}, SARA~\cite{xiao2019sara}, and Vanir~\cite{vanir} find near-optimal configurations for a group of workloads simultaneously.
Lynceus~\cite{casimiro2020lynceus} simultaneously optimizes the choice of configurations and application-level parameters such as the hyper-parameters of a machine learning algorithm.
Bilal et al.~\cite{bilal2020best} evaluate a wide range of black-box optimization algorithms and show that Bayesian optimization with gradient boosted regression trees performs well.
Google's Autopilot~\cite{rzadca2020autopilot} automatically scales the resource usage of workloads to reduce overallocation of resources while minimizing the risk of the application getting throttled or killed due to not having enough resources.

All these prior works focus on finding a near-optimal configurations that optimize a particular goal.
However, users may have diverse goals and preferences and some of these goals may be subtle.
For example, a user looking for the best-performing system and configuration may be willing to sacrifice a small amount of performance if this sacrifice can result in large cost savings.
Alternatively, a user who is optimizing for cost may be willing to incur slightly higher cost if it would result in substantially better performance.
For users to make such decisions (or employ automated tools for making such decisions), they need to be able to see the complete trade-off between performance and cost (or other metrics of interest) to select between different Pareto-optimal options.

A number of prior works aim at predicting the performance of applications on all available systems and configurations to provide users with a complete view of the trade-off space~\cite{ernest,baughman2018profiling,aaziz2018modeling,paris,xu2020arvmec,mariani2017predicting,paragon,quasar,lin2020modelling,chen2021silhouette}.
Some of these works use analytical models~\cite{ernest,baughman2018profiling,aaziz2018modeling} where performance is modelled with a hand-crafted equation and the coefficients of this equation are trained for each application and system.
For example, Ernest~\cite{ernest} proposes an equation with multiple terms that capture fixed costs, computations, and different communication patterns.
It uses non-negative least squares to train the coefficients of this equation, deriving different coefficients for every combination of application and type of system.
Baughman et al.~\cite{baughman2018profiling} propose an equation with different terms representing components of a configuration such as the amount of vCPUs or memory, and uses non-linear least squares to find coefficients for each application.
The drawback of these analytical approaches is that they tend to be specific to the application and/or system in question, which means that they require extensive online profiling of the application of interest on the system of interest before being able to make predictions for that application on that system.
Generalizing analytical models to work for multiple applications and systems would require considering a larger number of features that distinguish different applications and systems.
However, hand-crafting equations that take such a large number of features is a difficult task.
Instead, more sophisticated machine learning models can better handle such large feature sets.

A number of works propose general models that are trained using information collected offline about a wide variety of applications on a wide variety of configurations~\cite{paris,xu2020arvmec,mariani2017predicting,paragon,quasar,lin2020modelling,chen2021silhouette}.
Fu et al.~\cite{fu2021use} study the general applicability of machine learning for performance prediction.
PARIS~\cite{paris} trains a random forest model, and generates an application fingerprint to provide as an input to that model by profiling a user-provided short-running task that is representative of the application on a select set of configurations.
ARVMEC~\cite{xu2020arvmec} trains an XGBoost~\cite{chen2016xgboost} model, and uses a similar approach to PARIS for fingerprint (or gene) generation.
Mariani et al.~\cite{mariani2017predicting} train a random forest model, however the input to the model is a hardware-independent profile of the application that can be generated on any system but requires recompilation of the application.
Paragon~\cite{paragon} and Quasar~\cite{quasar} use collaborative filtering to predict the performance of an application across many systems and configurations, and use the predicted information for scheduling and resource assignment.
While these predictions are effective when used in such a fashion, prior work~\cite{paris} has found that the predictions provided by collaborative filtering do not have sufficient accuracy in their raw form and our experience has been the same.

Our proposed approach of generating an application fingerprint by profiling an application on select configurations then passing this fingerprint to a general model is inspired by the work in PARIS~\cite{paris}.
However, there are several aspects that distinguish our work from that in PARIS.
First, while PARIS predicts performance globally across multiple systems and configurations, we provide multiple scopes of prediction (global, single-system, and local) with varying accuracy and online profiling overhead catered to different use-cases.
Second, in PARIS, the fingerprint configurations are pre-determined or have to be selected by the deployer, whereas we provide an automated approach for identifying the best set of fingerprint configurations.
Third, PARIS requires users to submit a representative short-running task to generate a fingerprint, whereas we use the original application to generate the fingerprint and do not need to run it to completion.
Fourth, we use a classifier to distinguish applications that scale well from those that scale poorly and show that this proposed classifier is beneficial for performance.
Fifth, we predict the sensitivity of the application to different kinds of interference.

A number of works predict performance across multiple systems and/or configurations for specific applications such as distributed matrix-multiplication~\cite{son2018distributed,kim2019mpec}, deep learning training~\cite{dube2019ai}, Hadoop MapReduce jobs~\cite{herodotou2011no}, VASP jobs~\cite{wang2019novel}, and e-science workflows~\cite{nadeem2019using}.
Our work targets general applications and is not specific to a particular framework or application domain.

Many works aim to model the impact of interference on application execution~\cite{paragon,quasar,upredict,govindan2011cuanta,koh2007analysis}.
For example, Paragon~\cite{paragon} and Quasar~\cite{quasar} use collaborative filtering to predict how a new application would perform under different interference and heterogeneity conditions to be able to co-schedule applications effectively.
uPredict~\cite{upredict} uses microbenchmarks to train application-specific performance models that learn an application's sensitivity to the contention of different resources.
In our work, we place a bound on the sensitivity of an application to interference by predicting the application's performance under different interference extremes.

PseudoApp~\cite{tak2013pseudoapp} alleviates the need to migrate an application to the cloud to evaluate its performance by constructing a pseudo-application that mimics the original application's behavior.
Such an approach is complementary to our work, whereby a pseudo-application may be used in the fingerprinting process if the user wishes to see the performance-cost trade-off before migrating the application.

\section{Limitations and Future Work}

One limitation of our current work is that we have evaluated our tool only on data analytics and scientific computing benchmarks executing on single-node CPU systems.
Our future work involves expanding our evaluation to other classes of applications as well as other types of systems such as multi-node systems with network communication and systems with accelerators such as GPUs.
Another limitation is that while our tool achieves low prediction error for many benchmarks, the error for some benchmarks remains high.
Our future work involves extending the classification stage to screen out such benchmarks that are difficult to predict.
Another limitation is that our workflow currently fingerprints every submitted application to make a prediction for it.
However, one possible optimization is to maintain a database of fingerprints for applications that have been seen before to avoid fingerprinting the same application again.
The success of this optimization depends on the extent to which an application's profile and relative performance are dependent on the input dataset.
Investigating such an optimization is also part of our future work.

\section{Conclusion}\label{sec:conclusion}

We propose a tool for predicting the performance-cost trade-off of an application across different systems and different resource configurations per system.
Our tool profiles a submitted application on a small number of systems and configurations without needing to run it to completion to measure its execution time.
It then uses the collected profiling information to classify the application based on its scalability and predict the performance of the application on all systems and configurations.
It provides multiple scopes of prediction (global, single-system, and local) with varying accuracy and online profiling overhead, and it also predicts the sensitivity of applications to different types of interference.
We train our prediction models by profiling a large number of applications across a wide variety of configurations, and we automatically identify a good set of configurations for profiling a submitted application on when making predictions.
Our evaluation on three single-node CPU systems shows that our tool is effective at predicting the performance of applications across multiple systems and configurations with low error.

\section*{Acknowledgements}

This work was supported by Hewlett Packard Enterprise.
We thank the anonymous reviewers for their feedback which has helped improve the paper.
We also appreciate the feedback and assistance received from Aditya Dhakal, Alok Mishra, Barbara Chapman, Diman Zad Tootaghaj, Eitan Frachtenberg, Gourav Rattihalli, Lianjie Cao, Mohammad Sonji, Ninad Sanjay Hogade, Norbert Bianchin, Paulo Roberto Pereira de Souza Filho, Pedro Bruel, and Rolando Pablo Hong Enriquez.

\balance
\bibliographystyle{IEEEtran}
\bibliography{main}

\begin{thebibliography}{10}
\providecommand{\url}[1]{#1}
\csname url@samestyle\endcsname
\providecommand{\newblock}{\relax}
\providecommand{\bibinfo}[2]{#2}
\providecommand{\BIBentrySTDinterwordspacing}{\spaceskip=0pt\relax}
\providecommand{\BIBentryALTinterwordstretchfactor}{4}
\providecommand{\BIBentryALTinterwordspacing}{\spaceskip=\fontdimen2\font plus
\BIBentryALTinterwordstretchfactor\fontdimen3\font minus
  \fontdimen4\font\relax}
\providecommand{\BIBforeignlanguage}[2]{{%
\expandafter\ifx\csname l@#1\endcsname\relax
\typeout{** WARNING: IEEEtran.bst: No hyphenation pattern has been}%
\typeout{** loaded for the language `#1'. Using the pattern for}%
\typeout{** the default language instead.}%
\else
\language=\csname l@#1\endcsname
\fi
#2}}
\providecommand{\BIBdecl}{\relax}
\BIBdecl

\bibitem{cherrypick}
O.~Alipourfard, H.~H. Liu, J.~Chen, S.~Venkataraman, M.~Yu, and M.~Zhang,
  ``{CherryPick}: Adaptively unearthing the best cloud configurations for big
  data analytics,'' in \emph{14th USENIX Symposium on Networked Systems Design
  and Implementation (NSDI 17)}, 2017, pp. 469--482.

\bibitem{hsu2018arrow}
C.-J. Hsu, V.~Nair, V.~W. Freeh, and T.~Menzies, ``Arrow: Low-level augmented
  bayesian optimization for finding the best cloud {VM},'' in \emph{2018 IEEE
  38th International Conference on Distributed Computing Systems
  (ICDCS)}.\hskip 1em plus 0.5em minus 0.4em\relax IEEE, 2018, pp. 660--670.

\bibitem{liu2019accordia}
Y.~Liu, H.~Xu, and W.~C. Lau, ``Accordia: Adaptive cloud configuration
  optimization for recurring data-intensive applications,'' in
  \emph{Proceedings of the ACM Symposium on Cloud Computing}, 2019, pp.
  479--479.

\bibitem{klimovic2018selecta}
A.~Klimovic, H.~Litz, and C.~Kozyrakis, ``Selecta: Heterogeneous cloud storage
  configuration for data analytics,'' in \emph{2018 USENIX Annual Technical
  Conference (USENIX ATC 18)}, 2018, pp. 759--773.

\bibitem{hsu2018micky}
C.-J. Hsu, V.~Nair, T.~Menzies, and V.~Freeh, ``Micky: A cheaper alternative
  for selecting cloud instances,'' in \emph{2018 IEEE 11th International
  Conference on Cloud Computing (CLOUD)}.\hskip 1em plus 0.5em minus
  0.4em\relax IEEE, 2018, pp. 409--416.

\bibitem{xiao2019sara}
A.~Xiao, Z.~Lu, J.~Li, J.~Wu, and P.~C. Hung, ``{SARA}: Stably and quickly find
  optimal cloud configurations for heterogeneous big data workloads,''
  \emph{Applied Soft Computing}, vol.~85, p. 105759, 2019.

\bibitem{vanir}
M.~Bilal, M.~Canini, and R.~Rodrigues, ``Finding the right cloud configuration
  for analytics clusters,'' in \emph{Proceedings of the 11th ACM Symposium on
  Cloud Computing}, 2020, pp. 208--222.

\bibitem{casimiro2020lynceus}
M.~Casimiro, D.~Didona, P.~Romano, L.~Rodrigues, W.~Zwaenepoel, and D.~Garlan,
  ``Lynceus: Cost-efficient tuning and provisioning of data analytic jobs,'' in
  \emph{2020 IEEE 40th International Conference on Distributed Computing
  Systems (ICDCS)}.\hskip 1em plus 0.5em minus 0.4em\relax IEEE, 2020, pp.
  56--66.

\bibitem{bilal2020best}
M.~Bilal, M.~Serafini, M.~Canini, and R.~Rodrigues, ``Do the best cloud
  configurations grow on trees? {A}n experimental evaluation of black box
  algorithms for optimizing cloud workloads,'' \emph{Proceedings of the VLDB
  Endowment}, vol.~13, no.~12, pp. 2563--2575, 2020.

\bibitem{hsu2018scout}
C.-J. Hsu, V.~Nair, T.~Menzies, and V.~W. Freeh, ``Scout: An experienced guide
  to find the best cloud configuration,'' \emph{arXiv preprint
  arXiv:1803.01296}, 2018.

\bibitem{lin2020fife}
Y.~Lin, J.~Briggs, and A.~Barker, ``{FIFE}: an infrastructure-as-code based
  framework for evaluating {VM} instances from multiple clouds,'' in \emph{2020
  IEEE/ACM 13th International Conference on Utility and Cloud Computing
  (UCC)}.\hskip 1em plus 0.5em minus 0.4em\relax IEEE, 2020, pp. 91--100.

\bibitem{wu2021apollo}
Y.-W. Wu, Y.-J. Xu, H.~Wu, L.-G. Su, W.-B. Zhang, and H.~Zhong, ``Apollo:
  Rapidly picking the optimal cloud configurations for big data analytics using
  a data-driven approach,'' \emph{Journal of Computer Science and Technology},
  vol.~36, no.~5, pp. 1184--1199, 2021.

\bibitem{ernest}
S.~Venkataraman, Z.~Yang, M.~Franklin, B.~Recht, and I.~Stoica, ``Ernest:
  Efficient performance prediction for $\{$Large-Scale$\}$ advanced
  analytics,'' in \emph{13th USENIX Symposium on Networked Systems Design and
  Implementation (NSDI 16)}, 2016, pp. 363--378.

\bibitem{baughman2018profiling}
M.~Baughman, R.~Chard, L.~Ward, J.~Pitt, K.~Chard, and I.~Foster, ``Profiling
  and predicting application performance on the cloud,'' in \emph{11th IEEE/ACM
  International Conference on Utility and Cloud Computing (UCC)}, 2018.

\bibitem{aaziz2018modeling}
O.~Aaziz, J.~Cook, and M.~Tanash, ``Modeling expected application runtime for
  characterizing and assessing job performance,'' in \emph{2018 IEEE
  International Conference on Cluster Computing (CLUSTER)}.\hskip 1em plus
  0.5em minus 0.4em\relax IEEE, 2018, pp. 543--551.

\bibitem{paris}
N.~J. Yadwadkar, B.~Hariharan, J.~E. Gonzalez, B.~Smith, and R.~H. Katz,
  ``Selecting the best {VM} across multiple public clouds: A data-driven
  performance modeling approach,'' in \emph{Proceedings of the 2017 Symposium
  on Cloud Computing}, 2017, pp. 452--465.

\bibitem{xu2020arvmec}
Y.~Xu, J.~Li, Z.~Lu, J.~Wu, P.~C. Hung, and A.~Alelaiwi, ``{ARVMEC}: Adaptive
  recommendation of virtual machines for iot in edge--cloud environment,''
  \emph{Journal of Parallel and Distributed Computing}, vol. 141, pp. 23--34,
  2020.

\bibitem{mariani2017predicting}
G.~Mariani, A.~Anghel, R.~Jongerius, and G.~Dittmann, ``Predicting cloud
  performance for {HPC} applications: A user-oriented approach,'' in \emph{2017
  17th IEEE/ACM International Symposium on Cluster, Cloud and Grid Computing
  (CCGRID)}.\hskip 1em plus 0.5em minus 0.4em\relax IEEE, 2017, pp. 524--533.

\bibitem{paragon}
C.~Delimitrou and C.~Kozyrakis, ``Paragon: Qos-aware scheduling for
  heterogeneous datacenters,'' in \emph{Proceedings of the Eighteenth
  International Conference on Architectural Support for Programming Languages
  and Operating Systems}, ser. ASPLOS '13.\hskip 1em plus 0.5em minus
  0.4em\relax New York, NY, USA: Association for Computing Machinery, 2013, p.
  77–88.

\bibitem{quasar}
------, ``Quasar: Resource-efficient and qos-aware cluster management,'' in
  \emph{Proceedings of the 19th International Conference on Architectural
  Support for Programming Languages and Operating Systems}, ser. ASPLOS
  '14.\hskip 1em plus 0.5em minus 0.4em\relax New York, NY, USA: Association
  for Computing Machinery, 2014, p. 127–144.

\bibitem{lin2020modelling}
Y.~Lin, A.~Barker, and J.~Thomson, ``Modelling {VM} latent characteristics and
  predicting application performance using semi-supervised non-negative matrix
  factorization,'' in \emph{2020 IEEE 13th International Conference on Cloud
  Computing (CLOUD)}.\hskip 1em plus 0.5em minus 0.4em\relax IEEE, 2020, pp.
  470--474.

\bibitem{chen2021silhouette}
Y.~Chen, L.~Lin, B.~Li, Q.~Wang, and Q.~Zhang, ``Silhouette: Efficient cloud
  configuration exploration for large-scale analytics,'' \emph{IEEE
  Transactions on Parallel and Distributed Systems}, vol.~32, no.~8, pp.
  2049--2061, 2021.

\bibitem{van2005gromacs}
D.~Van Der~Spoel, E.~Lindahl, B.~Hess, G.~Groenhof, A.~E. Mark, and H.~J.
  Berendsen, ``{GROMACS}: fast, flexible, and free,'' \emph{Journal of
  computational chemistry}, vol.~26, no.~16, pp. 1701--1718, 2005.

\bibitem{chen2016xgboost}
T.~Chen and C.~Guestrin, ``Xgboost: A scalable tree boosting system,'' in
  \emph{Proceedings of the 22nd acm sigkdd international conference on
  knowledge discovery and data mining}, 2016, pp. 785--794.

\bibitem{docker}
\BIBentryALTinterwordspacing
S.~Hykes, ``Docker,'' 2022. [Online]. Available: \url{https://www.docker.com}
\BIBentrySTDinterwordspacing

\bibitem{perf}
A.~C. De~Melo, ``The new linux’perf’tools,'' in \emph{Slides from Linux
  Kongress}, vol.~18, 2010, pp. 1--42.

\bibitem{stress-ng}
\BIBentryALTinterwordspacing
C.~I. King, ``Stress-ng,'' 2022. [Online]. Available:
  \url{https://wiki.ubuntu.com/Kernel/Reference/stress-ng}
\BIBentrySTDinterwordspacing

\bibitem{bailey1991parallel}
D.~H. Bailey, E.~Barszcz, J.~T. Barton, D.~S. Browning, R.~L. Carter, L.~Dagum,
  R.~A. Fatoohi, P.~O. Frederickson, T.~A. Lasinski, R.~S. Schreiber
  \emph{et~al.}, ``The {NAS} parallel benchmarks summary and preliminary
  results,'' in \emph{Supercomputing'91: Proceedings of the 1991 ACM/IEEE
  Conference on Supercomputing}.\hskip 1em plus 0.5em minus 0.4em\relax IEEE,
  1991, pp. 158--165.

\bibitem{zhan2017parsec3}
X.~Zhan, Y.~Bao, C.~Bienia, and K.~Li, ``{PARSEC3.0}: A multicore benchmark
  suite with network stacks and {SPLASH-2X},'' \emph{ACM SIGARCH Computer
  Architecture News}, vol.~44, no.~5, pp. 1--16, 2017.

\bibitem{juckeland2014spec}
G.~Juckeland, W.~Brantley, S.~Chandrasekaran, B.~Chapman, S.~Che, M.~Colgrove,
  H.~Feng, A.~Grund, R.~Henschel, W.-M.~W. Hwu \emph{et~al.}, ``{SPEC ACCEL}: A
  standard application suite for measuring hardware accelerator performance,''
  in \emph{International Workshop on Performance Modeling, Benchmarking and
  Simulation of High Performance Computer Systems}.\hskip 1em plus 0.5em minus
  0.4em\relax Springer, 2014, pp. 46--67.

\bibitem{aslot2001specomp}
M.~S. M{\"u}ller, J.~Baron, W.~C. Brantley, H.~Feng, D.~Hackenberg,
  R.~Henschel, G.~Jost, D.~Molka, C.~Parrott, J.~Robichaux \emph{et~al.},
  ``{SPEC OMP2012} — an application benchmark suite for parallel systems
  using {OpenMP},'' in \emph{International Workshop on OpenMP}.\hskip 1em plus
  0.5em minus 0.4em\relax Springer, 2012, pp. 223--236.

\bibitem{stratton2012parboil}
J.~A. Stratton, C.~Rodrigues, I.-J. Sung, N.~Obeid, L.-W. Chang, N.~Anssari,
  G.~D. Liu, and W.-m.~W. Hwu, ``Parboil: A revised benchmark suite for
  scientific and commercial throughput computing,'' \emph{Center for Reliable
  and High-Performance Computing}, vol. 127, p.~27, 2012.

\bibitem{che2009rodinia}
S.~Che, M.~Boyer, J.~Meng, D.~Tarjan, J.~W. Sheaffer, S.-H. Lee, and
  K.~Skadron, ``Rodinia: A benchmark suite for heterogeneous computing,'' in
  \emph{2009 IEEE International Symposium on Workload Characterization
  (IISWC)}.\hskip 1em plus 0.5em minus 0.4em\relax IEEE, 2009, pp. 44--54.

\bibitem{meng2016mllib}
X.~Meng, J.~Bradley, B.~Yavuz, E.~Sparks, S.~Venkataraman, D.~Liu, J.~Freeman,
  D.~Tsai, M.~Amde, S.~Owen \emph{et~al.}, ``{MLlib: Machine learning in Apache
  Spark},'' \emph{The Journal of Machine Learning Research}, vol.~17, no.~1,
  pp. 1235--1241, 2016.

\bibitem{harris2020array}
\BIBentryALTinterwordspacing
C.~R. Harris, K.~J. Millman, S.~J. van~der Walt, R.~Gommers, P.~Virtanen,
  D.~Cournapeau, E.~Wieser, J.~Taylor, S.~Berg, N.~J. Smith, R.~Kern, M.~Picus,
  S.~Hoyer, M.~H. van Kerkwijk, M.~Brett, A.~Haldane, J.~F. del R{'{\i}}o,
  M.~Wiebe, P.~Peterson, P.~G{'{e}}rard-Marchant, K.~Sheppard, T.~Reddy,
  W.~Weckesser, H.~Abbasi, C.~Gohlke, and T.~E. Oliphant, ``Array programming
  with {NumPy},'' \emph{Nature}, vol. 585, no. 7825, pp. 357--362, Sep. 2020.
  [Online]. Available: \url{https://doi.org/10.1038/s41586-020-2649-2}
\BIBentrySTDinterwordspacing

\bibitem{reback2020pandas}
\BIBentryALTinterwordspacing
T.~pandas~development team, ``pandas-dev/pandas: Pandas,'' Feb. 2020. [Online].
  Available: \url{https://doi.org/10.5281/zenodo.3509134}
\BIBentrySTDinterwordspacing

\bibitem{pedregosa2011scikit}
F.~Pedregosa, G.~Varoquaux, A.~Gramfort, V.~Michel, B.~Thirion, O.~Grisel,
  M.~Blondel, P.~Prettenhofer, R.~Weiss, V.~Dubourg, and Others,
  ``{Scikit-learn: Machine learning in Python},'' \emph{the Journal of machine
  Learning research}, vol.~12, pp. 2825--2830, 2011.

\bibitem{smape}
\BIBentryALTinterwordspacing
Wikipedia, ``Symmetric mean absolute percentage error,'' 2022. [Online].
  Available:
  \url{https://en.wikipedia.org/wiki/Symmetric\_mean\_absolute\_percentage\_error}
\BIBentrySTDinterwordspacing

\bibitem{rzadca2020autopilot}
K.~Rzadca, P.~Findeisen, J.~Swiderski, P.~Zych, P.~Broniek, J.~Kusmierek,
  P.~Nowak, B.~Strack, P.~Witusowski, S.~Hand \emph{et~al.}, ``Autopilot:
  Workload autoscaling at {Google},'' in \emph{Proceedings of the Fifteenth
  European Conference on Computer Systems}, 2020, pp. 1--16.

\bibitem{fu2021use}
S.~Fu, S.~Gupta, R.~Mittal, and S.~Ratnasamy, ``On the use of {ML} for blackbox
  system performance prediction,'' in \emph{18th USENIX Symposium on Networked
  Systems Design and Implementation (NSDI 21)}, 2021, pp. 763--784.

\bibitem{son2018distributed}
M.~Son and K.~Lee, ``Distributed matrix multiplication performance estimator
  for machine learning jobs in cloud computing,'' in \emph{2018 IEEE 11th
  International Conference on Cloud Computing (CLOUD)}.\hskip 1em plus 0.5em
  minus 0.4em\relax IEEE, 2018, pp. 638--645.

\bibitem{kim2019mpec}
J.~Kim, M.~Son, and K.~Lee, ``{MPEC}: Distributed matrix multiplication
  performance modeling on a scale-out cloud environment for data mining jobs,''
  \emph{IEEE Transactions on Cloud Computing}, 2019.

\bibitem{dube2019ai}
P.~Dube, T.~Suk, and C.~Wang, ``{AI Gauge}: Runtime estimation for deep
  learning in the cloud,'' in \emph{2019 31st International Symposium on
  Computer Architecture and High Performance Computing (SBAC-PAD)}.\hskip 1em
  plus 0.5em minus 0.4em\relax IEEE, 2019, pp. 160--167.

\bibitem{herodotou2011no}
H.~Herodotou, F.~Dong, and S.~Babu, ``No one (cluster) size fits all: automatic
  cluster sizing for data-intensive analytics,'' in \emph{Proceedings of the
  2nd ACM Symposium on Cloud Computing}, 2011, pp. 1--14.

\bibitem{wang2019novel}
Q.~Wang, J.~Li, S.~Wang, and G.~Wu, ``A novel two-step job runtime estimation
  method based on input parameters in system,'' in \emph{2019 IEEE 4th
  International Conference on Cloud Computing and Big Data Analysis
  (ICCCBDA)}.\hskip 1em plus 0.5em minus 0.4em\relax IEEE, 2019, pp. 311--316.

\bibitem{nadeem2019using}
F.~Nadeem, D.~Alghazzawi, A.~Mashat, K.~Faqeeh, and A.~Almalaise, ``Using
  machine learning ensemble methods to predict execution time of e-science
  workflows in heterogeneous distributed systems,'' \emph{IEEE Access}, vol.~7,
  pp. 25\,138--25\,149, 2019.

\bibitem{upredict}
H.~Moradi, W.~Wang, A.~Fernandez, and D.~Zhu, ``{uPredict}: A user-level
  profiler-based predictive framework in multi-tenant clouds,'' in \emph{2020
  IEEE International Conference on Cloud Engineering (IC2E)}.\hskip 1em plus
  0.5em minus 0.4em\relax IEEE, 2020, pp. 73--82.

\bibitem{govindan2011cuanta}
S.~Govindan, J.~Liu, A.~Kansal, and A.~Sivasubramaniam, ``Cuanta: quantifying
  effects of shared on-chip resource interference for consolidated virtual
  machines,'' in \emph{Proceedings of the 2nd ACM Symposium on Cloud
  Computing}, 2011, pp. 1--14.

\bibitem{koh2007analysis}
Y.~Koh, R.~Knauerhase, P.~Brett, M.~Bowman, Z.~Wen, and C.~Pu, ``An analysis of
  performance interference effects in virtual environments,'' in \emph{2007
  IEEE International Symposium on Performance Analysis of Systems \&
  Software}.\hskip 1em plus 0.5em minus 0.4em\relax IEEE, 2007, pp. 200--209.

\bibitem{tak2013pseudoapp}
B.~C. Tak, C.~Tang, H.~Huang, and L.~Wang, ``{PseudoApp}: Performance
  prediction for application migration to cloud,'' in \emph{2013 IFIP/IEEE
  International Symposium on Integrated Network Management (IM 2013)}.\hskip
  1em plus 0.5em minus 0.4em\relax IEEE, 2013, pp. 303--310.

\end{thebibliography}

\end{document}